\documentclass[%
 reprint,
superscriptaddress,
amsmath,amssymb,amsthm,
aps,
]{revtex4-2}
\usepackage[normalem]{ulem}
\usepackage{graphicx}
\graphicspath{{./images/}}

\usepackage{amsthm}
\usepackage{hyperref}
\usepackage{enumitem}
\usepackage{xcolor}
\usepackage{lipsum}

\usepackage[linesnumbered,ruled,vlined]{algorithm2e}

\newtheorem{theorem}{Theorem}
\newtheorem*{theorem*}{Theorem}
\newtheorem{lemma}{Lemma}

\newtheorem{fact}{Fact}

\begin{document}

\title{High-performance local decoders for defect matching in 1D}

\author{Louis Paletta}
\email[]{louis.paletta@inria.fr}
\affiliation{Laboratoire de Physique de l'Ecole normale supérieure, ENS-PSL, CNRS, Inria, Centre Automatique et Systèmes (CAS), Mines Paris, Université PSL, Sorbonne Université, Université Paris Cité, Paris, France}
\author{Anthony Leverrier}
\affiliation{Inria Paris, France}
\author{Mazyar Mirrahimi}
\affiliation{Laboratoire de Physique de l'Ecole normale supérieure, ENS-PSL, CNRS, Inria, Centre Automatique et Systèmes (CAS), Mines Paris, Université PSL, Sorbonne Université, Université Paris Cité, Paris, France}
\author{Christophe Vuillot}
\thanks{Now at Alice\&Bob}
\affiliation{Universit\'e de Lorraine, CNRS, Inria, LORIA, F-54000 Nancy, France}

\date{\today}

\begin{abstract}
    Local decoders, also known as cellular-automaton decoders, offer a promising path toward real-time quantum error correction by replacing centralized classical decoding, with inherent hardware constraints, by a natively parallel and streamlined architecture from a simple local transition rule.
    We propose two new types of local decoders for the quantum repetition code in one dimension.
    The signal-rule decoders interpret odd parities between neighboring qubits as defects, attracted to each other via the exchange of classical point-like excitations, represented by a few bits of local memory.
    We prove the existence of a threshold in the code-capacity model and present numerical evidence of exponential logical error suppression under a phenomenological noise model, with data and measurement errors at each error correction cycle.
    Compared to previously known local decoders that suffer from sub-optimal threshold and scaling, our construction significantly narrows the gap with global decoders for practical system sizes and error rates.
    Implementation requirements can be further reduced by eliminating the need for local classical memories, with a new rule defined on two rows of qubits. This shearing-rule works well at relevant system sizes making it an appealing short-term solution. When combined with biased-noise qubits, such as cat qubits, these decoders enable a fully local quantum memory in one dimension.
\end{abstract}

\maketitle

Quantum error correcting codes protect information in a noisy quantum computer by delocalizing it across a higher-dimensional Hilbert space. 
Topological codes are a particularly efficient approach to realize a quantum memory where one enforces local constraints on physical qubits placed on a surface~\cite{kitaev2003fault}. They generally display a good resistance to noise, and the prototypical 2D surface code \cite{fowler2012surface} remains today the leading experimental platform \cite{google2023suppressing,krinner2022realizing,acharya2024quantum}.

In this setting, the error correction mechanism starts by measuring local stabilizers that give the \textit{syndrome}. A classical decoder then tries to identify the most likely error compatible with this syndrome. Efficient decoders exist for well-known topological codes \cite{edmonds1965paths,dennis2002topological,delfosse2021almost,chamberland2020triangular,kubica2023efficient,berent2024decoding}, but usually suffer from the following caveats: ($i$) the decoder requires access to the entire syndrome, which impose additional hardware capabilities in order to send all stabilizer measurements to a central computing unit, and ($ii$) extra redundancy in the syndrome is needed to cope with measurement errors, either by repeating the measurements in time, or by exploiting higher-dimensional codes~\cite{bombin2015single}. 
An attractive alternative to this active error correction strategy would be to rely on passive error correction via so-called \textit{self-correcting} quantum memories \cite{chesi2010self,yoshida2011feasibility,landon2015perturbative,brown2016quantum}, where information is encoded in the degenerate ground states of a Hamiltonian protected at finite temperature by a significant energy gap with the rest of the spectrum. The approach has been fruitful in the classical setting, where a bit of information can be passively stored in the macroscopic magnetization of a ferromagnetic material in dimension $\geq 2$ (Ising model), e.g.\ in hard-drives. While the 4D toric code offers a theoretical solution in the quantum case~\cite{alicki2010thermal,kitaev2003fault}, it is known that self-correcting quantum memories cannot exist in 2 dimensions \cite{bravyi2009no} and whether they can in 3 dimensions remains an outstanding open question in the field~\cite{bravyi2013quantum,lin2024proposals}.

\begin{table*}
    \begin{tabular}{llcl}
		\hline
		Decoder & Dim. & Threshold & Main idea \\ \hline
        Gács \cite{gacs2001reliable} & 1D & — & Hiercharical structure with constant-overhead self-simulation \\
		
		Tsirelson \cite{cirel2006reliable,balasubramanian2024local} & 1D* & 1.4\% & Large error clusters are recursively split into smaller, locally erasable clusters \\
		Harrington \cite{harrington2004analysis} & 1D* & 2.0\% & Local implementation of a renormalization-group-like decoder \\
		Field-based \cite{herold2015cellular,herold2017cellular} & 1D* & — & Local variables encode a classical field attracting defects to each other \\
		SSR [This work] & 1D* & 6.6\% & Defects attract each other by exchanging classical binary signals \\ 
        Toom's rule \cite{toom1995cellular} & 2D & 7.7\% & Localized error cluster are removed from their south-east corner \\ \hline
	\end{tabular}
	\caption[Local decoders of the repetition code]{Local decoders of the repetition code. The label 1D* refers to a one-dimensional lattice in which each site carries a logarithmic number of bits. The value of the threshold is given for phenomenological noise, where for Harrington's decoder defined initially for the toric code we define and simulate a repetition code variant for a fair comparison. Note that all listed decoders, at the exception of Gács' construction, are known to generalize to (or have been initially defined for) the toric code, at the cost of 'squaring the dimension', i.e.\ 4D toric code for a generalization of Toom's rule, and 2D toric code equipped with local logarithmic memories for 1D* decoders.}
	\label{table:local_decoders}
\end{table*}

Local decoders, also known as cellular-automaton decoders, offer a natural middle ground between global decoders and self-correcting quantum memories. In this case, the stabilizer measurement sites are equipped with a classical automaton that can perform a simple computation, communicate with its neighbors and apply a local correction on the state.
These actions form a local \textit{transition rule}, that can induce a non-trivial macroscopic dynamics useful for classical \cite{toom1980stable,gacs2001reliable,cirel2006reliable} and quantum \cite{kubica2019cellular,breuckmann2016local,harrington2004analysis,guedes2024quantum,brown2016quantum,vasmer2021cellular,ueno2021qecool} error correction purposes. A well-known example is \textit{Toom's rule} that protects classical information stored in a 2D grid for a time exponential in the grid size~\cite{toom1980stable}. This is achieved by emulating a digital version of the Ising model with a simple transition rule that flips a bit if it disagrees with both its West and South neighbors. 

\section{State of the art}
\subsection{Decoding the quantum repetition code}
We focus here on local decoders for the quantum repetition code under Pauli $X$ errors. While this is a very simple code, it can be combined with biased-noise qubits~\cite{leghtas2015confining,mirrahimi2014dynamically,puri2017engineering} to yield a quantum memory~\cite{guillaud2019repetition,ruiz2024ldpc,chamberland2022building,putterman2024hardware,lieu2024candidate}. In addition, it serves as a toy model for the surface code, but in one spatial dimension instead of two.

The $n$-qubit 1D repetition code is defined by placing qubits on the $n$ edges on a cycle and stabilizers $S_i := Z_{(i-1,i),(i,i+1)}$ on its vertices $i \in \mathbb{Z}_n$. It encodes a single logical qubit, with logical codewords $|k\rangle_L := |k\rangle^{\otimes n}$ for $k\in \{0,1\}$. The syndrome of an $X$-type error $E$ defined on the edges of the cycle corresponds to the boundary of the error. It will be convenient to represent it as $\Sigma = \partial E := \{\sigma_1, \ldots, \sigma_\ell\} \subseteq \mathbb{Z}_n$, i.e.\ as the set of vertices where the values of the incident edges differ. We take the convention that $\sigma_j < \sigma_{j+1}$ and remark that this set has even cardinality. The corresponding vertices carry point-like excitations that we call \textit{defects}.

The decoding problem is equivalent to finding a correct matching of these defects. There are only two solutions, corresponding to the true error $E$ and to its complement $\mathbb{Z}_n \setminus E$. The decoder succeeds if it correctly recovers $E$.
An optimal decoder is readily available if it has access to the full syndrome since it suffices to choose the matching that minimizes the weight of the error.
In addition to this \textit{code-capacity scenario} where each qubit belongs to $E$ independently with probability $\varepsilon$, we will also be concerned with the \textit{phenomenological error model} where each (data) qubit is flipped at each time step with probability $\varepsilon_d$ and each stabilizer measurement result is flipped with probability $\varepsilon_m$.

\subsection{Existing local decoders}
Self-correcting classical memories cannot exist in one spatial dimension. However, a seminal result of Gacs showed that a local decoder could display a phase-transition behaviour in the phenomenological model~\cite{gacs2001reliable}. His construction relies on a sophisticated cellular automaton with a hierarchical structure reminiscent of concatenated decoding. 
While a beautiful proof-of-principle, this decoder offers limited practical applications to error correction due to its high complexity and suboptimal performance. Here, we are concerned with the quantum setting and some assumptions relevant to the classical case can be relaxed: first it is standard to assume that classical computation is performed reliably \cite{dennis2002topological,gottesman2013fault}, and it is reasonable to allow the memory size to depend (logarithmically) on the system size. To distinguish such constructions from cellular automaton decoder, which usually implies a strictly constant state space, we will refer to such systems as local decoders.

\begin{figure*}
    \centering
    \includegraphics[width=\linewidth]{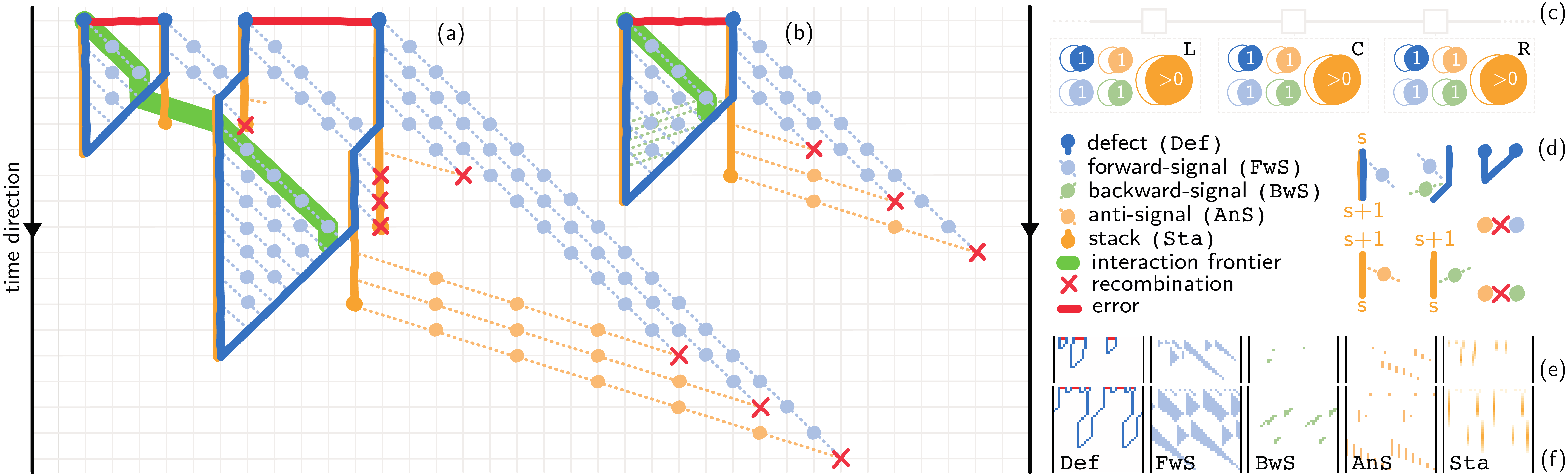}
    \caption{\label{figure:signal_rule} Illustration of the ASR decoder. (a) Simplified representation of the erasure of a complex cluster where anti-signals are only represented at their creation when on the left of the rightmost syndrome, and backward-signals are not represented. (b) Space-time representation of the erasure of a simple error cluster with time going downwards. Each defect sends forward-signals to its right until forward-signals sent from the left defect reaches the right defect, at which point the right defect is displaced to the left for each forward-signal it receives until it recombines with the left defect. Forward-signals that have induced a defect displacement transform into backward-signals that propagate in the opposite direction to recombine with the left stack. Forward-signals sent by the right defect recombine with faster anti-signals created from the decrement of the right stack when it no longer coincides with a defect. (c) Local representation of the decoder variables. (d) Summary of excitations creation and annihilation rules with the value of the stack indicated by $s \geq 0$. (e) is a complete space-time representation of the corrected error configuration illustrated in (a-b) and (f) of a fractal-like error configuration of weight $w<n/2$ resulting in a logical error. Binary variables are represented with the same colors as in (a-d) while we use a color gradient for the stack variable.}
\end{figure*}

Proposals for local decoders fall into two main categories:
decoders with a hierarchical structure \cite{balasubramanian2024local,harrington2004analysis,breuckmann2016local} inspired by classical constructions \cite{cirel2006reliable,gacs2001reliable}, and field-based decoders \cite{herold2015cellular,herold2017cellular,michnicki2015towards} where defects are interpreted as particles interacting with each other through a classical field simulated by the classical automaton. 
Hierarchical constructions \cite{balasubramanian2024local,harrington2004analysis,breuckmann2016local} protect information for a time $\exp(\gamma_n)$ with $\gamma_n \propto n^\alpha$ for $\alpha>0$, but suffer from a low-error threshold for $\varepsilon_d$ and $\varepsilon_m$, and from a poor \textit{effective distance} $\gamma_n$, corresponding to the minimal weight of error configuration leading to a logical failure. Recall that the distance of the repetition code is $n$, and the effective distance therefore quantifies how well the local decoder performs compared to a global decoder.  
Field-based decoders \cite{herold2015cellular,herold2017cellular,michnicki2015towards,guedes2024quantum,lake2025fast}, on the other hand, display high thresholds and good performance for small system size but the memory lifetime saturates above a certain system size, unless one keeps on increasing the communication speed. Existing proposals are summarized in Table~\ref{table:local_decoders}. While these decoders are typically designed for the surface code, they are either directly inspired from a construction for the repetition code in 1D, or can be easily be adapted to that case, with a similar expected qualitative behavior. The converse is not necessarily true, in particular if the decoder takes specifically advantage of the 1D structure or relies on qubit permutations that leave the stabilizers of the repetition code invariant, but not those of the surface code.

\section{Main results}
In this paper, we define two new 1D-local decoders for the quantum repetition code, the \textit{asymmetric signal-rule} (ASR) and the \textit{symmetric signal-rule} (SSR). The ASR is the simpler one to describe and analyze, but its symmetrized version, the SSR, leads to better performance.
We compare the decoders with a variant of Toom's rule on a flat 2D surface and with a cellular automaton that we call \textit{shearing rule} which induces a dynamics similar to that of the recent \textit{quantum two-line voting} scheme~\cite{guedes2024quantum,lang2018strictly}, albeit with a significantly reduced footprint. The shearing rule appears to work well for systems of small size and is described in Section~\ref{section:shearing}.

We first prove a threshold theorem for the ASR in the code-capacity model, when the rule is applied on each site for $\tau = \mathcal{O}(n)$ time steps. 
\begin{theorem}[ASR code-capacity threshold]\label{theorem:threshold_code_capacity}
    There exist $\varepsilon_{th}>0, \alpha >0$ and $\tau = \mathcal{O}(n)$ such that for $\varepsilon < \varepsilon_{\text{th}}$, the logical error rate $\varepsilon_L$ of the ASR applied for $\tau$ time steps to an initial error where each qubit is flipped independently and identically with probability $\varepsilon$ satisfies $\varepsilon_L \leq \exp(-n^{\alpha})$.
\end{theorem}
We show that $\alpha>0.12$ in the proof, but our numerical simulations (in the more general phenomenological model) indicate that this bound is far from tight. The proof combines the fact that an isolated cluster of errors is erased in a time proportional to its size, with a general technique of hierarchical decomposition of error configurations from renormalization group decoders \cite{bravyi2013quantum,harrington2004analysis,gacs2001reliable,kubica2019cellular}. A cluster of errors belonging to a given level of the hierarchy is erased independently from upper levels so that only error clusters from sufficiently high levels can induce a logical flip. The proof of erasure of an isolated is outlined in Section~\ref{section:signal-rule_decoders}, and included in details alongside the hierarchical decomposition in the Appendix.

Then we numerically evaluate the performance of SSR for the phenomenological model, as illustrated in Figure~\ref{figure:performances}, and pick $\varepsilon_d = \varepsilon_m = \varepsilon$ unless stated otherwise. We find that the SSR approaches the performance of global decoders in terms of effective distance for small system sizes, and we show evidence for asymptotic exponential suppression of the logical error probability in $n^\alpha$ with $\alpha > 0$. In practice, the 1D SSR outperforms Toom's rule for practically relevant parameters, without exhibiting any saturation of the logical error rate for large $n$.

\section{Signal-rule decoders}\label{section:signal-rule_decoders}
\subsection{Asymmetric signal-rule}
The local decoder is defined via a transition rule that updates classical variables assigned to each vertex of $\mathbb{Z}_n$. Each such site hosts four binary registers encoding the presence of four types of point-like particles: defects, forward-signals, backward-signals and anti-signals, as well as an additional stack register serving as a reservoir of anti-signals. A site is represented on Figure~\ref{figure:signal_rule} (c) and a configuration $u^{(t)} \in U_n := (\mathbb{Z}_2^4 \times \mathbb{N})^n$ of the decoder at time $t$ corresponds to the value of all variables on all sites.
The decoder mediates an attractive interaction between defects through the exchange of signals. We sketch its dynamics here. At each iteration, a defect emits a forward-signal, which propagates to the right until meeting another defect. In that case, the encountered defect moves one step to the left and the forward-signal becomes a backward-signal traveling more rapidly to the left. To keep track of the forward-signals that have been emitted, one increments the local stack at the defect site when a forward-signal is sent, and decrements it when a backward-signal comes back, in which case the backward-signal is also annihilated. In addition, when a stack is no longer associated with a defect (because the latter has moved to the left), it sends anti-signals that move quickly to the right and whose goal is to recombine with the remaining forward or backward-signals. A sequence of decoder configuration $u^{(t)} \in U_n$ is represented on Figure~\ref{figure:signal_rule} (a-b).

\begin{figure*}
    \centering
    \includegraphics[width=\linewidth]{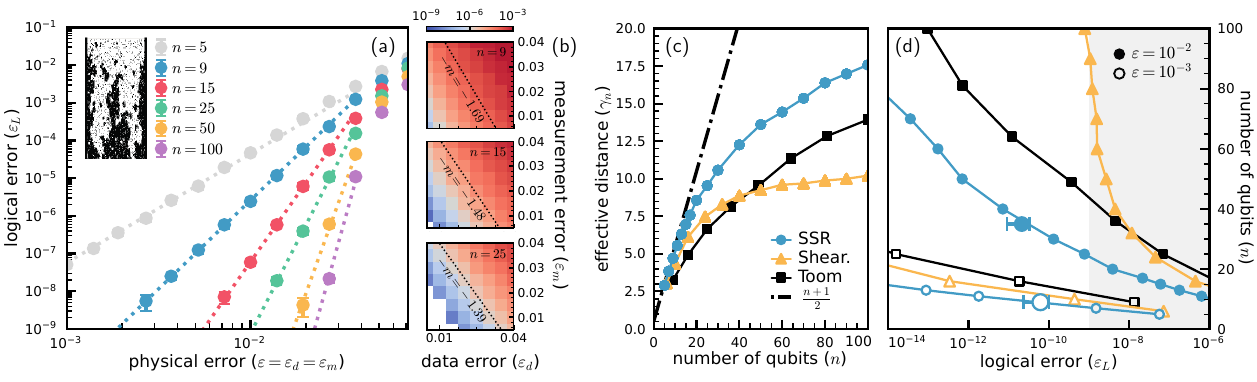}
    \caption{\label{figure:performances} Performances of the SSR decoder. (a) Logical error rate as a function of the physical error rate for several system sizes in the phenomenological model. The logical error rate is obtained from the normalization of the logical flip rate estimated on Monte Carlo simulations for a fixed time $\tau$ large enough so that $\varepsilon_L$ is independent of $\tau$. The data is fitted with the ansatz $An(B\varepsilon)^{\gamma_n}$ with a different parameter $\gamma_n$ for each $n$. $B^{-1}$ is estimated to $6.6\%$. (Inset) Space-time representation of a logical bit-flip for $n=100$, $\tau = 200$ and $\varepsilon=6\%$, with time flowing downwards. (b) Logical error rate as a function of the data and measurement error probabilities $\varepsilon_d$ and $\varepsilon_m$. We fit $\varepsilon_L$ to the ansatz $A_n(\varepsilon_d+\varepsilon_m/m)^{\gamma_n}$ where $\gamma_n$ is taken from (a) and draw an associated contour line for $n=9,15,25$. (c) Effective distance $\gamma_n$ as a function of $n$ for the SSR, the shearing-rule and Toom's rule. The dashed line represents the scaling of the repetition code in $\frac{n+1}{2}$ when decoded using minimum-weight perfect matching \cite{edmonds1965paths,dennis2002topological}. (d) Estimate of the number of physical qubits necessary to reach a given logical error rate based on the ansatz fitted on the grey region. Larger points with error bars in the white region are obtained numerically and confirm the validity of the estimates in that regime.
    The lines in the last two plots are for visual guidance only and do not represent actual data.}
\end{figure*}


Algorithm \ref{algorithm:ASR} describes the update rule on a central site \texttt{C}. The automaton can read the values of the registers at its neighboring left and right sites, \texttt{L} and \texttt{R}, but can only update the registers at $\texttt{C}$. Each step of the rule is applied in parallel on all the sites. 
The state of the automaton at the beginning of an iteration on site $\texttt{X}\in \{\texttt{L},\texttt{C},\texttt{R}\}$ is given by the values \texttt{Var.X} for $\texttt{Var}\in \{ \texttt{Def}, \texttt{FwS}, \texttt{BwS}, \texttt{AnS},\texttt{Sta}\}$. Within an iteration we use the temporary variables \texttt{Tmp} and \texttt{Cor} for information transfer between sites and to save the correction. Note that in the phenomenological model, the value of \texttt{Def.C} is reset to the parity measurement between the left and right edges at the beginning of each iteration, and we apply an $X$-type correction to the left edge if $\texttt{Cor.C}=1$ at the end.
Anti-signals and backwards-signals travel respectively at speed $k_a \geq 3$ and $k_b \geq 2$. We set $k_a=k_b=3$ in the following and in numerical simulations.
Finally, the symmetric signal-rule is obtained by combining an ASR as above, pointing to the right, with a second ASR pointing to the left so that the attraction works in both directions, as detailed at the end of this section.

\medbreak

\begin{algorithm}\label{algorithm:ASR}
    \SetKwBlock{RepeatKa}{repeat $k_a-1 \ times$}{}
    \SetKwBlock{RepeatKb}{repeat $k_b \ times$}{}
    \caption{ASR update rule}\label{algorithm:SSR}
    \KwIn{Local variables \texttt{Var.X} for \texttt{Var} in \{\texttt{Def}, \texttt{FwS}, \texttt{BwS}, \texttt{AnS}, \texttt{Sta}\} and \texttt{X} in \{\texttt{L}, \texttt{C}, \texttt{R}\}}
    \KwOut{Local variables}

    \medbreak
    
    Initialize \texttt{Cor.C} and \texttt{Tmp.C} to 0\;

    \emph{Matching of neighboring defects}\;
    \If{$(\mathtt{Def.L}, \mathtt{Def.C}, \mathtt{Def.R}) = (0,1,1)$}{
        Set \texttt{Cor.C} to 1; set \texttt{Def.C} to 0\;
    }
    \If{$\mathtt{Cor.L}=1$}{
        Set \texttt{Def.C} to 0
    }

    \emph{Send forward-signals}\;
    \If{$(\mathtt{Def.C}, \mathtt{Def.R}) = (1,0)$ \textnormal{\textbf{and}} $\mathtt{FwS.C} = 0$}{
        Set \texttt{FwS.C} to 1; increment \texttt{Sta.C} by 1\;
    }
    \emph{Propagate forward-signals to the right}\;
    Set \texttt{Tmp.C} to \texttt{FwS.C}; \texttt{FwS.C} to \texttt{Tmp.L}; and \texttt{Tmp.C} to 0 \;

    \emph{Correction and signals reflection}\;
    \If{$\mathtt{Def.C} = \mathtt{FwS.C} = 1$}{
         Set \texttt{Tmp.C} to 1; set \texttt{Def.C} to 0; set (\texttt{FwS.C}, \texttt{BwS.C}) to $(0,1)$ \textbf{if} $(\texttt{FwS.C}, \texttt{BwS.C}) = (1,0)$
    }
    \If{$\mathtt{Tmp.R} = 1$}{
        Set \texttt{Cor.C} to 1; set \texttt{Def.C} to 1; set (\texttt{FwS.C}, \texttt{BwS.C}) to $(0,1)$ \textbf{if} $(\texttt{FwS.C}, \texttt{BwS.C}) = (1,0)$
    }
    \emph{Propagate backward-signals by $k_b$ to the left and recombine with anti-signals and stack}\;
    \RepeatKb{
        Set \texttt{Tmp.C} to \texttt{BwS.C} and \texttt{BwS.C} to \texttt{Tmp.R}\;
        \textnormal{\textbf{if}} $\texttt{BwS.C}=\texttt{AnS.C}=1$ \textnormal{\textbf{then}} set both to 0\;
        \textnormal{\textbf{if}} $\texttt{BwS.C}=1$ \textnormal{\textbf{and}} $\texttt{Sta.C}>0$ \textnormal{\textbf{then}} set $\texttt{BwS.C}$ to 0 and decrement $\texttt{Sta.C}$ by 1\;
    }

    \emph{Send anti-signals}\;
    \If{$\mathtt{Def.C} = \mathtt{AnS.C} = 0$ \textnormal{\textbf{and}} $\mathtt{Sta.C} > 0$}{
        Set \texttt{AnS.C} to 1; decrement \texttt{Sta.C} by 1\;
    }
    \emph{Propagate anti-signals by $k_a$ to the right and recombine with forward and backward-signals}\;
    \RepeatKa{
        Set \texttt{Tmp.C} to \texttt{AnS.C} and \texttt{AnS.C} to \texttt{Tmp.L}\;
        \textnormal{\textbf{if}} $\texttt{AnS.C}=\texttt{FwS.C}=1$ \textnormal{\textbf{then}} set both to 0\;
        \textnormal{\textbf{if}} $\texttt{AnS.C}=\texttt{BwS.C}=1$ \textnormal{\textbf{then}} set both to 0\;
    }
    Set \texttt{Tmp.C} to \texttt{AnS.C} and \texttt{AnS.C} to \texttt{Tmp.L}\;
    \textnormal{\textbf{if}} $\texttt{AnS.C}=\texttt{BwS.C}=1$ \textnormal{\textbf{then}} set both to 0\;
\end{algorithm}

\subsection{Erasure of a finite-size error}

The main step in the proof of Theorem \ref{theorem:threshold_code_capacity} is to show that on an infinite lattice $\mathbb{Z}$, the ASR erases any finite-size error $E \subset \mathbb{Z}$ of width $\Delta$ in $\mathcal{O}(\Delta)$ time steps. The case of a connected error pattern, illustrated on Figure~\ref{figure:signal_rule} (b), is the simplest one. The initial defects are located at sites $\sigma_1$ and $\sigma_2 = \sigma_1 + \Delta$. The first forward-signal emitted from $\sigma_1$ reaches $\sigma_2$ at time $t=\Delta$, the defects recombine at time $t=2\Delta$, and the choice $k_a\geq 2$ ensures that all forward-signals have been caught by anti-signals at time $t \leq 3\Delta$, guaranteeing that the decoder has reached the zero-error configuration.

In general, an error may consist of $m \geq 1$ clusters, giving rise to $m$ pairs of defects $\Sigma = \{\sigma_1, ..., \sigma_{2m} \} \subset \mathbb{Z}$ with $\sigma_1 < ... < \sigma_{2m}$. These defects can interact via the exchange of signals, leading to a complex dynamics, as depicted on Figure~\ref{figure:signal_rule} (a).
We show that all defects and excitations (either signals or stack increments) eventually recombine in a time linear in the width $\Delta :=\sigma_{2m} - \sigma_1$ of the error. Note that the leftmost defect $\sigma_1$ never moves when working on an infinite lattice.
We give a brief outline of the proof here, with the full proof deferred to the Appendix. The proof works by a reduction to the single cluster case. The main idea is to define space-time \textit{interaction frontier} (depicted in green in Figure~\ref{figure:signal_rule} and defined in the Appendix) such that below this frontier, the odd-numbered defects remain immobile and recombine with their right neighbor, as if there was only a single pair of defects. We show that the frontier reaches the last defect in time $\mathcal{O}(\Delta)$.

It is left to prove that after all defects have recombined, all excitations recombine too in time linear in the width $\Delta$ of the error so that the decoder converges correctly to the zero configuration. This follows from a conservation law between excitations: we assign a $+1$ charge to every forward-signal and backward-signal, and a $-1$ charge to every anti-signal and stack increment, while the defects don't carry any charge. We prove that the total charge within the decoder is conserved (a corollary of which the stack on a given site is bounded by $2n$ on a finite size lattice) and that each positive charge can be paired up with a negative charge on its left. Since stack increments eventually transforms into anti-signals in the absence of defects, choosing $k_a \geq 2$ ensures that the anti-signals will catch forward-signals, and that all excitations will eventually recombine.


\subsection{Symmetric signal-rule}\label{subsection:ssr}

The symmetric signal-rule (SSR) is obtained by combining an ASR pointing to the right, with a second ASR pointing to the left so that the attraction works in both directions. This comes at the cost of doubling the number of variables. We indicate by 1 and 2 the ASR pointing respectively to the right and to the left so that the state of the automaton at the beginning of an iteration on site $\texttt{X}\in \{\texttt{L},\texttt{C},\texttt{R}\}$ is given by the values \texttt{Var.i.X} for $\texttt{Var}\in \{ \texttt{Def}, \texttt{FwS}, \texttt{BwS}, \texttt{AnS},\texttt{Sta}\}$ and $\texttt{i}\in \{1,2\}$. 
The variables of each ASR evolve simultaneously and independently according to Algorithm~\ref{algorithm:ASR}, with
the exception of the two correction steps described in the algorithms below. The first one corresponds to the recombination of neighboring defects while the second one deals with displacing defect receiving forward-signals. In the two cases the modification aims at building agreement between the two ASR to avoid correcting twice the same error.

\renewcommand{\thealgocf}{2a}
\begin{algorithm}\label{algorithm:SSR_1}\caption{SSR update rule: \textit{Matching of neighboring defects}}
    \If{$(\mathtt{Def.1.L}, \mathtt{Def.1.C}, \mathtt{Def.1.R}) = (0,1,1)$}{
        Set \texttt{Tmp.1.C} to 1\;
    }
    \If{$(\mathtt{Def.2.L}, \mathtt{Def.2.C}, \mathtt{Def.2.R}) = (1,1,0)$}{
        Set \texttt{Tmp.2.C} to 1\;
    }
    \If{$\mathtt{Tmp.1.C}=1$ \textnormal{\textbf{or}} $\mathtt{Tmp.2.R}=1$}{
        Set \texttt{Cor.C} to 1; set \texttt{Def.1.C} and \texttt{Def.2.C} to 0\;
    }
    \For{$\mathtt{(i,X)} \in \{(\mathtt{1},\mathtt{L}),(\mathtt{2},\mathtt{R})\}$}{
        \If{$\mathtt{Cor.X} = 1$}{
            Set \texttt{Def.i.C} to 0\;
    }
    }
    Set \texttt{Tmp.1.C} and \texttt{Tmp.2.C} to 0\;
\end{algorithm}

\renewcommand{\thealgocf}{2b}
\begin{algorithm}\label{algorithm:SSR_2}\caption{SSR update rule: \textit{Correction and signals reflection}}
    \For{$\mathtt{i} \in \{1,2\}$}{
        \If{$\mathtt{Def.i.C} = \mathtt{FwS.i.C} = 1$}{
            Set \texttt{Tmp.i.C} to 1\;
            Set $(\mathtt{FwS.i.C}, \mathtt{BwS.i.C})$ to $(0,1)$ 
            \textbf{if} $(\mathtt{FwS.i.C}, \mathtt{BwS.i.C}) = (1,0)$\;
        }
    }
    \If{$\mathtt{Tmp.1.C} + \mathtt{Tmp.2.R} = 1$}{
        Set \texttt{Cor.C} to 1; set \texttt{Def.1.C} and \texttt{Def.2.C} to 0\;
    }
    \For{$\mathtt{(i,X)} \in \{(\mathtt{1},\mathtt{L}),(\mathtt{2},\mathtt{R})\}$}{
        \If{$\mathtt{Cor.X} = 1$}{
        Set \texttt{Def.i.C} to 1\;
        Set (\texttt{FwS.i.C}, \texttt{BwS.i.C}) to $(0,1)$ \textbf{if} $(\mathtt{FwS.i.C}, \mathtt{BwS.i.C}) = (1,0)$\;
    }
    }
\end{algorithm}

\section{Dynamics under phenomenological noise}

\subsection{Markovian dynamics}\label{appendix:markov}

\begin{figure}
    \centering
    \includegraphics[width=\linewidth]{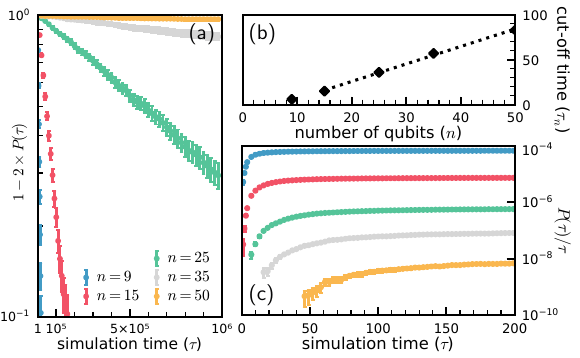}
    \caption{\label{figure:markov} Markovian dynamics. (a) $1-2P_{L}(\tau)$, as a function of $\tau$, a constant slope in logarithmic scale indicates a logical flip probability independent of the time of the simulation. The logical error rate $\varepsilon_L$ is estimated in the asymptotic regime where $P_{L}(\tau)/\tau$ is a constant. (b) Average convergence time to this asymptotic regime, defined as $\tau_n := \min\{\tau_0 \geq 0 \mid P_{L}(\tau)/\tau > \varepsilon_L/2, \tau \geq \tau_0 \}$. We fit $\tau_n$ with the linear ansatz $an+b$ and we find $a=1.95$. (c) $P_{L}(\tau)/\tau$ as a function of $\tau$ for different system sizes, used to compute $\tau_n$.
    }
\end{figure}

We evaluate the performance of the decoder under phenomenological noise by numerical simulations for $t \leq \tau$ for some fixed $\tau \geq 0$. In usual decoding schemes, the information from a stabilizer measurement is erased after some time, usually linear in the distance of the code, evacuating the entropy from the system. This is not the case with the ASR and SSR decoders where a forward-signal, i.e.\ a previous odd parity measurement, a priori does not have a definite lifetime. This makes the Markovian behavior of the decoder non-trivial while this property is required in order to define a logical error rate. We confirm this behavior numerically in the following.

The decoder is initialized in the zero configuration and we denote by $P_L(\tau)$ the probability of a logical flip (in this case defined by a majority of $1$) at time $\tau$. We observe numerically in Figure~\ref{figure:markov} (a) that $P_L(\tau)$ follows asymptotically a Poisson-like behavior, i.e.\ of the form $P_L(\tau) \simeq [1-(1-\varepsilon_L)^\tau]/2$ for some $\varepsilon_L > 0$ defining the logical error rate. Note that in the absence of anti-signals ensuring a balance between excitations, the occurrence rate of logical failure  would typically increase with time as signals accumulate within the system.

The logical failure of the decoder results in the general case from space-time error clusters. Hence the simulation time should be large enough to account for the contribution of the most likely error clusters resulting in a logical flip. This numerically corresponds to the convergence of the normalized logical flip probability $P_L(\tau)/\tau$ towards the logical error rate, which is shown in Figure~\ref{figure:markov} (c). We define the convergence time $\tau_n$ for $n$ qubits to be the minimum time after which $P_L(\tau)/\tau > \varepsilon_L/2$, and we show $\tau_n$ as a function of $n$ in Figure~\ref{figure:markov} (b). Numerical fits give $\tau_n = \mathcal{O}(n)$, and at the exception of high logical error rates close to the threshold, numerical simulations use $\tau = 1000$ which is one order of magnitude greater than any expected convergence time. Note that since the logical error rate is normalized, longer simulation time does not reduce numerical performances.

\subsection{Performance}
We plot the logical error rate as a function of the physical error rate and $n$ in Figure~\ref{figure:performances} (a), and as a function of the data error and measurement error probabilities $\varepsilon_d$ and $\varepsilon_m$ in Figure~\ref{figure:performances} (b). We observe in the latter that the logical error rate depends on a linear function of $\varepsilon_d$ and $\varepsilon_m$ which confirms the intuition that, in a local decoding scheme, measurement errors and data errors play similar roles.

We compare the SSR decoder with a variant of Toom’s rule defined on a flat 2D surface, and with a new cellular automaton decoder we introduce in Section~\ref{section:shearing} and that we call the \textit{shearing-rule}. The latter decoder is defined on two concentric cycles of qubits, and induces a similar dynamics to that of the quantum implementation \cite{guedes2024quantum} of the two-line voting decoder \cite{toom1995cellular,gacs1978one}, albeit with significantly reduced quantum circuit complexity, making it of interest on its own. The SSR decoder experiences fractal-like error configurations leading to logical failure. This is also the case of the ASR, and we present an example for the more restricted code-capacity model in Figure~\ref{figure:signal_rule} (f). Such configurations are responsible for the sublinear scaling of the effective distance and are characteristic of renormalization-group decoders \cite{rozendaal2023worst}. 
We assess the practical performance of the various decoders by fitting $\varepsilon_L$ with an ansatz of the form $An(B\varepsilon)^{\gamma_n}$, where the exponent $\gamma_n$ is allowed to depend on $n$: see Figure~\ref{figure:performances} (c). For the SSR decoder, we obtain $A = 2.1 \times 10^{-3}$ and $B^{-1} = 6.6\%$. Note that comparing $\gamma_n$ to $(n+1)/2$ helps assess the performance loss relative to global decoders like minimum-weight perfect matching that decode up to the optimal distance. The behavior of $\gamma_n$ as a function of $n$ for the SSR decoder provides evidence of the exponential suppression of the logical error rate with increasing system size in the asymptotic regime. In contrast, $\gamma_n$ saturates from $n \simeq 30$ onward for the shearing-rule. For small system sizes, however, numerical simulations show that the SSR and shearing-rule decoders exhibit similar performance, both outperforming Toom’s rule.

\subsection{Resources requirements}
Every global decoder can be implemented as a local decoder at the cost of considerable local classical memories. We show here that only a few bits on each decoder site is required for the ASR and SSR decoders to work. The ASR requires one binary register for each of the four types of point-like excitations, a register able to store an integer to represent the stack, typically on a logarithmic number of bits with binary encoding, and a few auxiliary bits for information transfer and local computation. This material cost simply doubles for the SSR, and we numerically evaluate below the size of the stack register that is needed in practice.

\begin{figure}
    \centering
    \includegraphics[width=\linewidth]{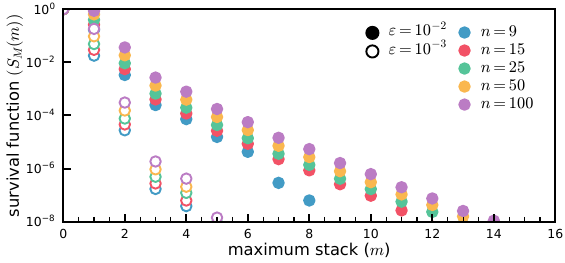}
    \caption{Classical resources requirements: survival function of the maximum stack height $M$ of a configuration upon successive SSR applications in the phenomenological model. $S_M(m) := \mathbb{P}(M \geq m)$ is estimated from Monte-Carlo simulations, and shown for various system sizes and physical error rates. Binary encoding of the stack further reduces the practical memory requirements to $\log m$. \label{figure:stack}}
\end{figure}

Consider a sequence of configurations in the phenomenological model upon application of the SSR decoder. Let $M^{(t)} \leq 2n$ be the random variable corresponding to the maximum value held in any of the $2n$ stacks (ASR pointing to the right or left, for each site) at time $t \geq 0$. For a Markovian dynamics, $M^{(t)}$ quickly converges to the time-independent associated random variable $M$. We estimate its survival function $S_M(m) := \mathbb{P}(M \geq m)$ from $M^{(t)}$ for $t \leq \tau = 1000$ on $\geq 10^6$ trajectories, and where $\tau$ is large compared to the convergence time of the normalized logical flip probability, see Figure~\ref{figure:markov}. We plot a numerical estimate of $S_M(m)$ for different physical error rates and system sizes in Figure~\ref{figure:stack}, and we observe that $S_M(m)$ is exponentially suppressed in $m$ in the regime $m = \mathcal{O}(n)$. The variable $S_M(m)$ can be used to estimate the maximum value that a stack should be able to encode to achieve a given logical error rate. For example, in the worst case considered in the simulations ($n=100$ and $\varepsilon = 10^{-2}$), the SSR decoder only requires stack with values $\leq 14$ to reach a logical error rate of $10^{-8}$. This can be done on 4 bits for each of the two stacks of a site.

\section{Eliminating the need for a classical memory}\label{section:shearing}

\subsection{The shearing-rule}

Here, we introduce a new cellular automaton that exhibits dynamics similar to those described in \cite{guedes2024quantum} but with reduced circuit practical complexity resulting in improved performances. This cellular automaton performs well for small system sizes and offers a practical advantage over ASR and SSR by eliminating the need for classical memory. As a result, it presents an appealing short-term solution for the local decoding of the repetition code.

\begin{figure}
    \centering
    \includegraphics[width=\linewidth]{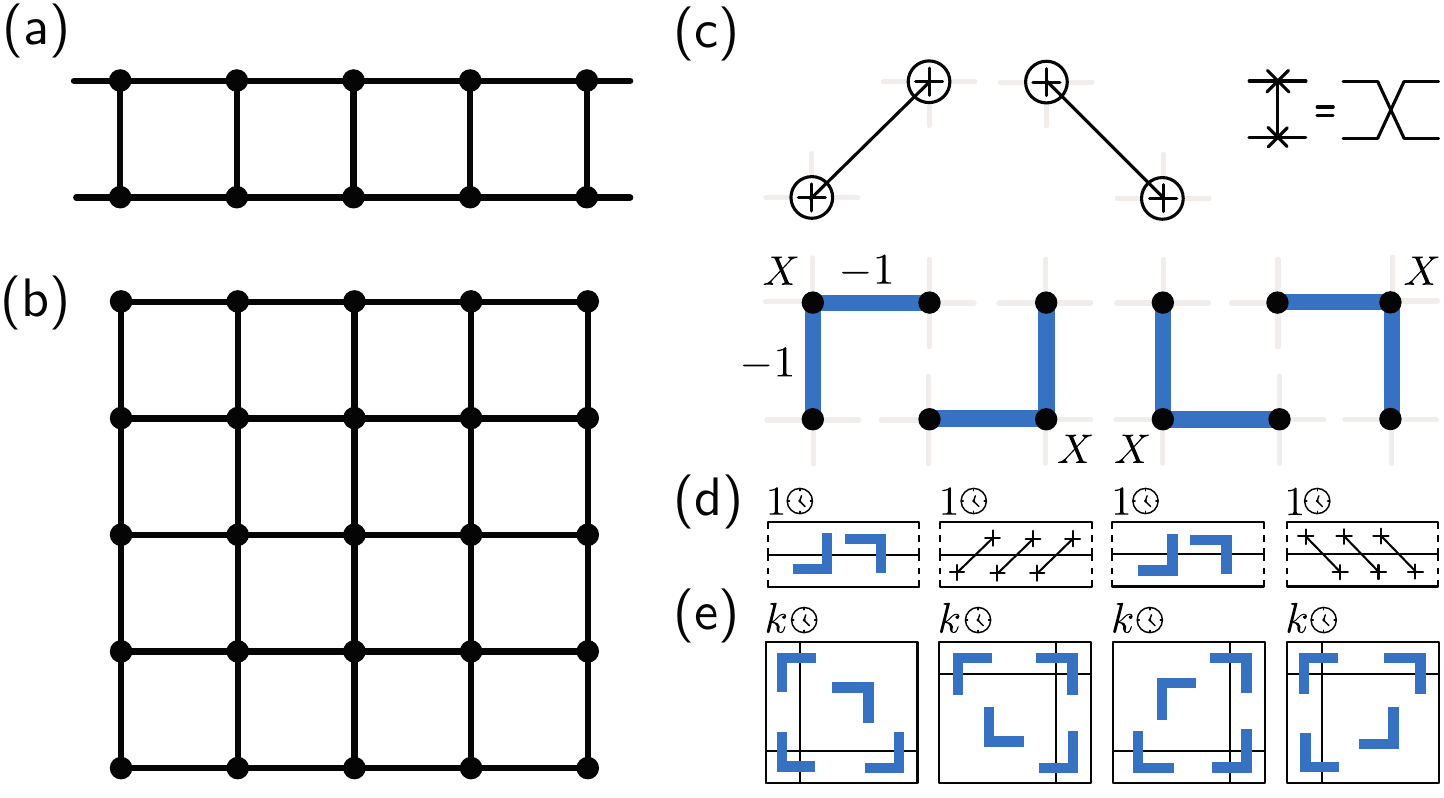}
    \caption{The shearing-rule and Toom's rule. (a-b) Representation of the two-row periodic lattice and the square lattice, on which are defined the shearing rule and Toom's rule, respectively. Qubits are assigned to vertices, and $Z$-type stabilizers to edges. (c) Elementary operations used in the two automata: qubit permutations and three qubits majority votes where blue edges accounts for odd parity. (d-e) Schematic representation of the temporal pattern of the shearing and Toom's rule where the colors indicate the orientation of the three qubits majority vote. Each update rule is applied once for the shearing rule, and $k = \Theta(\log n)$ times for Toom's rule. \label{figure:shearing_toom}}
\end{figure}

Let $n$ be an even integer. Consider the periodic lattice $\mathbb{Z}_{2} \times \mathbb{Z}_{n/2}$ and assign respectively to its vertices and edges, qubits and $Z$-type stabilizers (parity check), i.e.\ the converse of the assignment used for the 1D repetition code. The $Z$-type stabilizers for $i < n/2$ are of the following types, $S_{0,i} := Z_{0,i}Z_{0,i+1}$, $S_{1,i} := Z_{1,i}Z_{1,i+1}$ and $S_{i} := Z_{0,i}Z_{1,i}$ and generate the $n$-qubit repetition code stabilizer group. The shearing-rule is defined on $\mathbb{Z}_{2} \times \mathbb{Z}_{n/2}$, with following elementary operations illustrated in Figure~\ref{figure:shearing_toom}: ($i$) $X$-type correction for qubits of the top row (resp. bottom) in odd parity with their bottom (resp. top) and left neighbours, ($ii$) qubits permutation along the left diagonal and ($iii$) qubits permutation along the right diagonal. The cellular automaton consists of the repeated sequential application of ($i$), ($ii$), ($i$), ($iii$) in that order. Note that the orientation of the majority vote of step ($i$) can be changed on the bottom row to make the error correction work on a non-periodic lattice. Up to a row permutation, we note that the cellular automaton induces a similar dynamics to that of \cite{guedes2024quantum}, where a cluster over two rows is separated by the dynamics into two separate clusters respectively on the top and bottom row. At this point the two clusters are erased independently by the cellular automaton decoder.

Note that for the repetition code, the permutation of two qubits can be performed via a SWAP gate but also by a parity measurement followed by an $XX$ correction conditioned on a $-1$ outcome. Concerning majority votes, a straightforward circuit implementation uses a Toffoli-gate per majority vote. The Toffoli count is crucial because, in practical physical implementations, three-qubit gates are likely to introduce the most errors. Notably, our construction requires only one-third as many Toffoli gates per site per correction step compared to \cite{guedes2024quantum}. 

Similarly as for the SSR, the logical performance of the cellular automaton decoder is investigated numerically in the phenomenological noise model 
with measurement and data errors for step ($i$) and data errors at steps $(ii)$ and $(iii)$ (that are without measurement). We plot the effective distance in Figure~\ref{figure:performances} (c), which saturates from $n \simeq 30$ onward similarly to \cite{guedes2024quantum}. This is due to the existence of logarithmic weight space-time error configuration leading to a logical failure. A simple example corresponds to an initial rectangular error cluster, for which the size is multiplied by some $\eta>1$ in the presence of an adversarial pair of errors (one on the left, one on the right), before the cluster separates in two.

\subsection{Toom's rule on a square lattice}

The two local decoders are compared with an implementation of Toom's rule \cite{toom1995cellular} on a 2D square lattice. The initial construction is defined on a torus and so we define boundary conditions in Figure~\ref{figure:shearing_toom}. In order to avoid errors to accumulate in a corner, the orientation of the majority vote in the bulk of the lattice is changed at time intervals which are logarithmic in the system size, similarly as in \cite{vasmer2021cellular}, the boundary conditions are updated accordingly.

\section{Discussion}

In this paper, we have introduced two new local decoders, the symmetric and asymmetric signal-rules, which can serve as a decoder for the quantum repetition code. We analytically prove the existence of a nonzero threshold in the code-capacity model for the asymmetric variant and provide evidence of stability for the symmetric variant when errors occur at each time step.
Immediate applications include the construction of a one-dimensional quantum memory obtained by concatenating a quantum repetition code, protected against bit-flips by a signal-rule decoder, with a bosonic code tailored for protection against phase-flips, e.g.\ cat codes \cite{mirrahimi2014dynamically,puri2017engineering,leghtas2015confining}, in the spirit of \cite{lieu2024candidate}.

Given the large number of possible signal-rule variants, we do not claim optimality of the decoder in terms of either performance or resource efficiency. The charge conservation property of signal-rules decoders however gives them a high-degree of structure, a desirable property enabling to reduce the space size in which to look for better performing variants. While, to the best of our knowledge, charge conservation in cellular automata has not been studied extensively, a related construction is that of number-conserving automata that conserves a number of particles~\cite{wolf2004lattice,nagel1992cellular}, with existing characterization in low dimensions~\cite{hattori1991additive,redeker2022number,wolnik2020split,boccara1998cellular,boccara2002number}. Extending the framework of number-conserving cellular automata to incorporate signed quantities (i.e., charges rather than particle counts) would prove particularly useful to look for charge-conserving decoder with good performances. Alternatively, a more general but promising approach to optimize the local update rule involves reinforcement learning~\cite{park2024enhancing}. Overall, we believe there remains significant room for performance improvement, which we leave for future work.

\section*{Code availability}
The code used for numerical simulations of the local decoders, analysis and visualization is available here \cite{git}.

\section*{Acknowledgments}
The authors thank Alain Sarlette for valuable discussions. LP thanks Diego Ruiz for stimulating discussions. We acknowledge support from the Plan France 2030 through the project NISQ2LSQ ANR-22-PETQ- 0006, HQI ANR-22-PNCQ-0002. Part of this research was performed while the author was visiting the Institute for Mathematical and Statistical Innovation (IMSI), which is supported by the National Science Foundation (Grant No. DMS-1929348).

\appendix

\clearpage

\section{Asymmetric signal-rule on an infinite lattice}\label{appendix:erasure}

The main step in the proof of Theorem~\ref{theorem:threshold_code_capacity} (proven in Section~\ref{appendix:proof_of_threshold}) is to show that on an infinite lattice $\mathbb{Z}$, the ASR erases any finite-size error $E \subset \mathbb{Z}$ of width $\Delta$ in $\mathcal{O}(\Delta)$ time steps. Let $u \in U := (\mathbb{Z}_2^4\times \mathbb{N})^{\mathbb{Z}}$ be a configuration of the decoder, and $\Sigma^{(t)} := \Sigma(u^{(t)})$ be the set of defects of the configuration of the decoder at time $t \geq 0$. We establish the following erasure theorem:

\begin{theorem}[Linear Erasure]\label{theorem:erasure_s}
    Let \( m \geq 1 \) and \(\Sigma := \{\sigma_1, \dots, \sigma_{2m} \} \ \mathrm{with} \ \sigma_1 < \dots < \sigma_{2m}\). Let $\Delta := \sigma_{2m}-\sigma_1$ and \( u^{(t)} \) initialized with $\Sigma^{(0)} = \Sigma$ and all other variables set to zero. We have for all \( t \geq 0 \),
    \begin{equation}
        \Sigma^{(t)} \subset [\sigma_1, \sigma_1 + \Delta], \quad  
        \mathrm{supp}(u^{(t)}) \subset [\sigma_1, \sigma_1 + 78\Delta].
    \end{equation}
    and for all $t > 77\Delta$,
    \begin{equation}
        \mathrm{supp}(u^{(t)}) = \varnothing.
    \end{equation}
\end{theorem}

Importantly, the decoder returns to the zero configuration after the initial error is corrected, that is to say all excitations have recombined. We introduce the \textit{interaction frontier} in subsection~\ref{appendix:interaction_frontier} that is central to the proof of the theorem, provided in~\ref{appendix:proof_theorem_erasure}. We establish some properties of the interaction frontier in \ref{appendix:independent_matching} and \ref{appendix:from_one_defect} that prove that the initial error is eventually corrected. It is left to ensure that all excitations recombine which follows from the properties of the charge distribution within the decoder that are proven in \ref{appendix:recombination}.

\subsection{Interaction frontier}\label{appendix:interaction_frontier}

The proof uses the non-decreasing space-time \textit{interaction frontier} $\varphi^{(t)} \in \mathbb{Z}$ defined for $t \leq t_r$, where $t_r$ will be specified later. Below this frontier, the odd defects remain immobile and recombine with their right neighbor, as if there was only a single pair of defects. We introduce the set of forward signals (denoted $\Phi_F^{(t)} := \Phi_F(u^{(t)})$) and the set of negative charges, i.e.\, the locations of non-zero stacks or anti-signals (denoted $\Phi_N^{(t)}$), both defined analogously to $\Sigma^{(t)}$ for defects. $\varphi^{(t)}$ is defined recursively from $\varphi^{(0)} = \mathrm{min}(\Sigma)$ by

\begin{equation}
    \varphi^{(t+1)} =
    \begin{cases}
        \varphi^{(t)} + 1, & \text{if } A \\
        \varphi^{(t)}, & \text{if } \neg A \land B \\
        \min ((\Sigma^{(t+1)} \cup \Phi_F^{(t+1)} )\cap \mathbb{Z}_{> \varphi}^{(t)}), & \text{if } \neg A \land \neg B
    \end{cases}
\end{equation}

where we use the notations $\mathbb{Z}_{> \varphi}^{(t)} = (\varphi^{(t)},+\infty)$ and

\begin{align*}
    A &: \varphi^{(t)} + 1 \in \Sigma^{(t+1)} \cup \Phi_F^{(t+1)} \\
    B &: |\Sigma^{(t)} \cap \mathbb{Z}_{\leq \varphi}^{(t)}| \equiv 1 \pmod{2}
\end{align*}

Here, \textit{A} means the right neighbor of the frontier hosts either a defect or a forward signal, while \textit{B} means there is an odd number of defects on the left of the frontier. The definition remains valid until some time $t_r \geq 0$ such that $\Sigma^{(t_r + 1)} \cap \mathbb{Z}_{> \varphi^{(t_r)}} = \varnothing$, i.e.\ until there are no defects remaining on the right of the frontier. Intuitively, the interaction frontier follows the propagation of a forward-signal until the said forward-signal recombine or reaches a defect. In such case, the forward-signal is attached to a defect or another forward-signal depending on the parity of the number of defects on the left of the frontier. The interaction frontier is illustrated in Figure~\ref{figure:signal_rule}. Note that in the figure we represent independently two interaction frontiers for two error clusters as if they were on separate lattices.

\subsection{Proof of erasure}\label{appendix:proof_theorem_erasure}

\begin{figure*}
    \centering
    \includegraphics[width=0.95\linewidth]{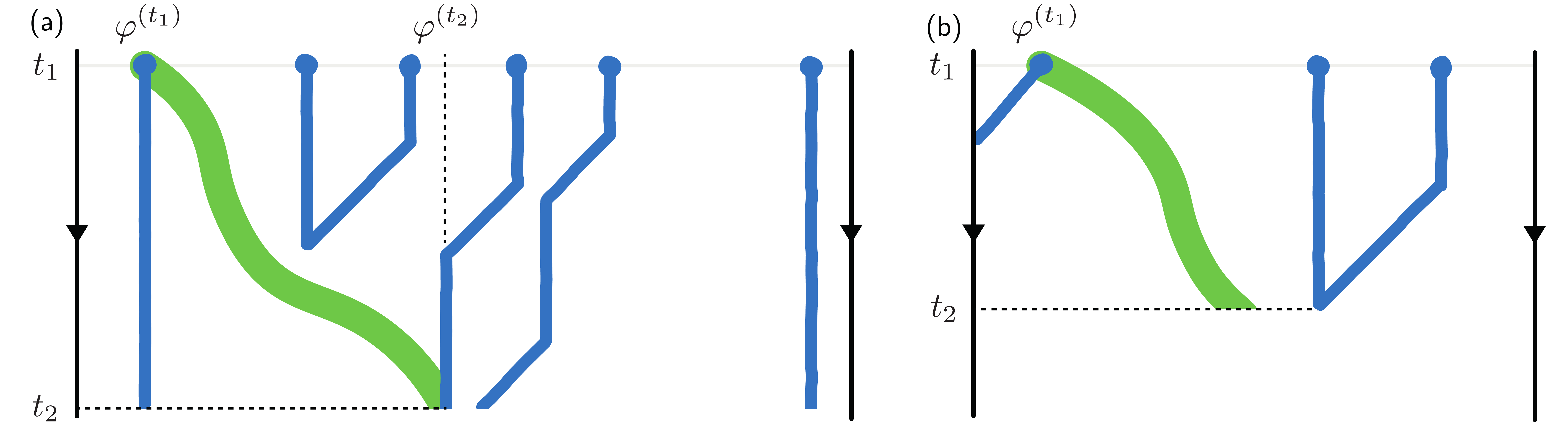}
    \caption{\label{figure:lemma_s1} Illustration of the two possible cases of Lemma~\ref{lemma:ifp_1_s} that characterize the increase of the interaction frontier depicted in green. (a) The interaction frontier goes from $\varphi^{(t_1)} \in \Sigma^{(t_1)}$ to $\varphi^{(t_2)} \in \Sigma^{(t_2)}$ in time $t_2-t_1 \leq 11(\varphi^{(t_2)}-\varphi^{(t_1)})$. (b) At time $t_2$ we have $\Sigma^{(t_2)} \cap \mathbb{Z}_{>\varphi}^{(t_2)} = \varnothing$. The repeated application of Lemma~\ref{lemma:ifp_1_s} implies that at some point all defects are on the left of the interaction frontier.}
\end{figure*}

\begin{figure*}
    \centering
    \includegraphics[width=0.7\linewidth]{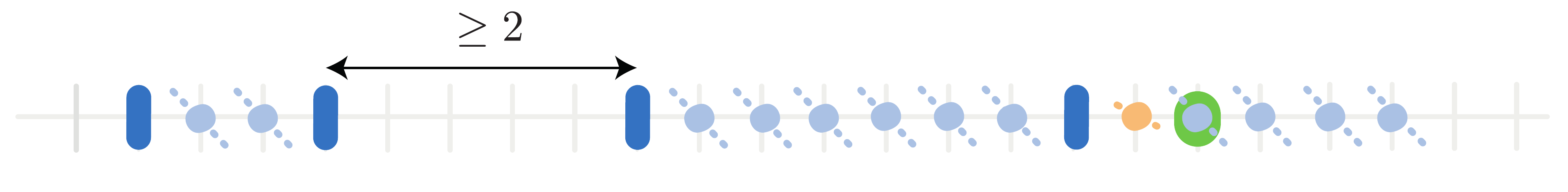}
    \caption{\label{figure:lemma_s2} Illustration of Lemma~\ref{lemma:ifp_2_s}, which characterizes the region to the left of the interaction frontier (shown in green). The open interval between an even and the subsequent odd defect is filled with forward signals and contains no negative charges. In contrast, the open interval between an odd and the following even defect has length at least 2 and contains no forward signals. This structure ensures that even defects are isolated from interference originating on the left, allowing defect pairs to recombine independently.}
\end{figure*}

\begin{figure*}
    \centering
    \includegraphics[width=0.65\linewidth]{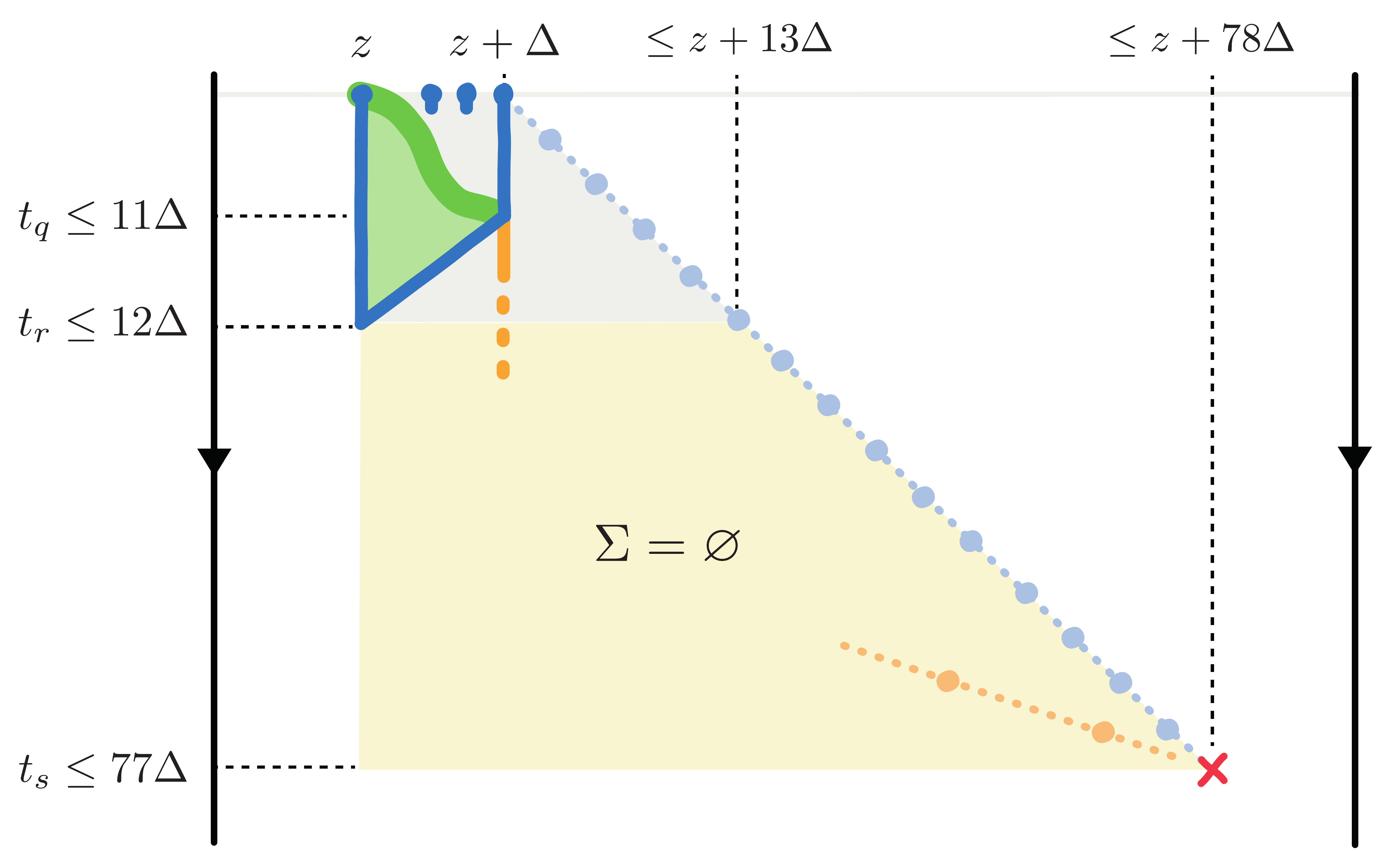}
    \caption{\label{figure:demo_th2} Illustration of the proof of Theorem~\ref{theorem:erasure_s} (not to scale). Let $\Sigma$ be a set of defects with even cardinality and diameter $\Delta$. By applying Lemma~\ref{lemma:ifp_1_s}, we show that the interaction frontier reaches the final defect by time $t_q \leq 11\Delta$. At this point, the region to the left of the frontier (depicted in light green) satisfies the conditions of Lemma~\ref{lemma:ifp_2_s}, ensuring that all defects recombine by time $t_r \leq 12\Delta$. The subsequent evolution of the decoder (yellow area), after defect recombination, is governed by Lemma~\ref{lemma:ifp_3_s}, which bounds the time of the last excitation recombination by $t_s \leq 77\Delta$. In the worst case, the final forward signal to recombine may have propagated a distance of at most $77\Delta$ from position $z + \Delta$, i.e., up to $z + 78\Delta$.}
\end{figure*}

We start by a brief outline of the proof. We show that the interaction frontier reaches the last defect in a time linear in the initial error width (using Lemma~\ref{lemma:ifp_1_s}, illustrated in Figure\ref{figure:lemma_s1} and proven in subsection~\ref{appendix:from_one_defect}). At this point all defects are on the left of the interaction frontier. This region is characterized so that pairs of defect recombine independently (Lemma~\ref{lemma:ifp_2_s}, illustrated in Figure~\ref{figure:lemma_s2} and proven in subsection~\ref{appendix:independent_matching}). It is left to prove that the dynamics of the decoder induces the recombination of all remaining excitations (Lemma~\ref{lemma:ifp_3_s}, illustrated in Fig~\ref{figure:demo_th2} proven in subsection~\ref{appendix:recombination}).

Before formally stating the relevant lemmas, we introduce some additional notations. Let $\Sigma_1$ and $\Sigma_2$ denote the sets of odd and even defects, respectively. We consider the two possible unions of integer open intervals formed between the even and odd defects, depending on whether the odd defects are positioned on the left or the right.

\begin{equation}
    \Sigma_{k,k+1} := \bigcup_{\substack{1 \leq i \leq 2m-1 \\ i \equiv k \, (\text{mod } 2)}} (\sigma_i, \sigma_{i+1}).
\end{equation}

We adopt the abbreviated notation $\Sigma_{12} := \Sigma_{1,2}$ and $\Sigma_{21} := \Sigma_{2,3}$ hereafter. Note that the integer interval $[\min(\Sigma), \max(\Sigma)]$ is partitioned as follows: $[\min(\Sigma), \max(\Sigma)]=\Sigma_1 \sqcup \Sigma_2 \sqcup \Sigma_{12} \sqcup \Sigma_{21}$. All these notations naturally extend to incorporate time dependence. With these definitions in place, we are now prepared to introduce the key lemmas required for the proof of Theorem~\ref{theorem:erasure_s}.

\begin{lemma}[From one defect to another]\label{lemma:ifp_1_s}
    Let $t_1 \geq 0$ such that $\varphi^{(t_1)} \in \Sigma^{(t_1)}$ and $\Sigma^{(t_1)} \cap \mathbb{Z}_{>\varphi}^{(t_1)} \neq \varnothing$, there exists $t_2 \in (t_1,t_1 + 3(\sigma_{2m}-\varphi^{(t_1)}))$ such that $\Sigma^{(t_2)} \cap \mathbb{Z}_{>\varphi}^{(t_2)} = \varnothing$ or $\varphi^{(t_2)} \in \Sigma^{(t_2)}$ with $\varphi^{(t_2)}-\varphi^{(t_1)} \geq (t_2-t_1)/11$.
\end{lemma}

Lemma~\ref{lemma:ifp_1_s} states that the interaction frontier exhibits a coarse-grained nonzero velocity: although the frontier may remain stationary over several iterations, its average speed over time is lower-bounded by $1/11$. This guarantees that the interaction frontier will eventually reach the right-most defect.

\begin{lemma}[Independent matching]\label{lemma:ifp_2_s}
    Let $t\geq 0$ and $\varphi^{(t)} \in \mathbb{Z}$ be the interaction frontier, the following holds
    \begin{align}
        &\big(\Sigma^{(t)}_{12} \cap \mathbb{Z}_{\leq \varphi}^{(t)}\big) \subset \Phi_F^{(t)}, \label{eq:lemma2_1}\\
        &\big(\Sigma^{(t)}_{12} \cap \Phi_N^{(t)} \cap \mathbb{Z}_{\leq \varphi}^{(t)}\big) = \varnothing, \label{eq:lemma2_2}\\
        &\big(\Sigma^{(t)}_{21} \cap \Phi_F^{(t)} \cap \mathbb{Z}_{< \varphi}^{(t)}\big) = \varnothing. \label{eq:lemma2_3}
    \end{align}
    In addition, for all $\sigma_a \in \Sigma_2^{(t)}, \sigma_b \in \Sigma_1^{(t)}$ such that $\sigma_a < \sigma_b \leq \varphi^{(t)}$, we have
    \begin{equation}
        \sigma_b-\sigma_a \geq 2. \label{eq:lemma2_4}
    \end{equation}
\end{lemma}

Here we characterize the region on the left of the interaction frontier (see Figure~\ref{figure:lemma_s2}) so that the interval between successive even and odd defects if filled with forward-signals, while the converse type of interval is without forward-signals and of length at least 2. This ensures that even and odd defects on the left of the frontier recombine independently.

\begin{lemma}[Excitations recombination]\label{lemma:ifp_3_s}
    Let $t_r \geq 0$, $z \in \mathbb{Z}$ and $\delta \geq 0$ such that
    \begin{equation}
        \Sigma^{(t_r)}=\varnothing \ \mathrm{and} \ \mathrm{supp}(u^{(t_r)}) \subset [z,z+\delta]
    \end{equation}
    then for all $t \geq t_r$ we have
    \begin{equation}
        \mathrm{supp}(u^{(t)}) \subset [z,z+6\delta]
    \end{equation}
    and for all $t \geq t_r +5\delta$ we have
    \begin{equation}
        \mathrm{supp}(u^{(t)}) = \varnothing.
    \end{equation}
\end{lemma}

Lemma~\ref{lemma:ifp_3_s} states that once all defects have recombined, all excitations eventually recombine in linear time with distance between the left-most and the right-most excitation at that time. In this sense, the correction of an error cluster occurs through a build-up of excitations within the decoder, which later recombine to restore the system to its initial zero configuration.

\medbreak

\begin{proof}[Proof of Theorem~\ref{theorem:erasure_s}] The proof is illustrated in Figure~\ref{figure:demo_th2}. Lemma~\ref{lemma:ifp_1_s} implies that there exists $t_q > 0$ such that $\Sigma^{(t_q)} \subset \mathbb{Z}_{\leq \varphi}^{(t_q)}$. In order to see this, we denote by $\{t_i\}_{ i\geq 1}$ with $t_1=0$ the strictly increasing set of times such that $\varphi^{(t_i)} \in \Sigma^{(t_i)}$. The application of Lemma \ref{lemma:ifp_1_s} for each $t_i$ gives
\begin{align}
    t_k = & \sum_{i=1}^{k-1} (t_{i+1}-t_i) \\
    \leq & 11\sum_{i=1}^{k-1} (\varphi^{(t_{i+1})}-\varphi^{(t_{i})})  \leq 11\Delta.
\end{align}
This upper bound implies that the sequence necessarily stops for a large enough $k=q-1$. With Lemma~\ref{lemma:ifp_1_s}, this means that there exists $t_q \leq 11\Delta$ such that $\Sigma^{(t_{q})}=\varnothing$ or $\Sigma^{(t_q)}\cap\mathbb{Z}_{>\varphi}^{(t_{q})}=\varnothing$.

At this point, all defects are either removed, or are to the left of the interaction frontier. By Lemma~\ref{lemma:ifp_2_s}, the open interval between odd and even defects is filled with forward-signals, while the interval between even and odd defects has length at least 2 and contains no forward-signals. Consequently, each odd defect remains stationary, while the corresponding even defect moves leftward at each iteration until they recombine. This ensures that the system becomes defect-free at time $t_r = t_q + \Delta$. Finally, the rightmost part of the support of $u^{(t)}$ is bounded by the position of a forward-signal emitted from $\sigma_{2m}$ at the first iteration, which has propagated for $t_r$ steps. Applying Lemma~\ref{lemma:ifp_3_s} with $z = \sigma_1$ and $\delta = t_r + \Delta = 13\Delta$ completes the proof of Theorem~\ref{theorem:erasure_s}.
\end{proof}

\subsection{Charge properties}\label{appendix:charge_properties}

We start by analyzing the dynamics of the decoder in terms of charged particles, which will be useful in the rest of the proof. We assign a $+1$ charge to every forward-signal and backward-signal, and a $-1$ charge to every anti-signal and stack increment, while the defects don't carry any charge. Let $q_i(u)$ be the total charge on site $i \in \mathbb{Z}$ of configuration $u \in U$. We define $Q(u)$ and $Q_z(u)$ the total charge respectively on all sites of configuration $u$ and on sites $< z \in \mathbb{Z}$. We use the following notation to account for time dependency

\begin{align}\label{eq:definition_conservation}
    &Q^{(t)} := Q(u^{(t)}) = \sum_{i \in \mathbb{Z}} q_i(u^{(t)}), \notag \\ &Q_z^{(t)} := Q_z(u^{(t)}) = \sum_{i < z} q_i(u^{(t)}).
\end{align}

We summarize the key charge properties of the decoder in three Facts stated below, with proof given at the end of the subsection.

\begin{fact}[Charge deficit]\label{fact:conservation}
    For all $t \geq 0$ we have
    \begin{equation}
        Q^{(t)} = 0, \label{eq:global}
    \end{equation}
    and for all $z \in \mathbb{Z}$,
    \begin{equation}
        Q_z^{(t)} \leq 0. \label{eq:deficit}
    \end{equation}
\end{fact}

Fact~\ref{fact:conservation} states that the global charge is conserved within the decoder, and that every positive charge can be paired up with a corresponding negative charge on its left. Note that the former property is also true on a finite lattice of size $n$, a corollary of which is that the stack variable on a given site is upper bounded by $2n$.

\begin{fact}[No recombination across defects]\label{fact:reco_accross}
    Let $t \geq 0$ and $\sigma_a^{(t)} < \sigma_b^{(t)} \in \Sigma^{(t)}$ be a pair of successive defects. No charge at time $t$ in the interval $[\sigma_a^{(t)}, \sigma_b^{(t)})$ recombines with a charge from $\mathbb{Z} \setminus [\sigma_a^{(t)}, \sigma_b^{(t)})$ during iteration $t$.
\end{fact}

Fact~\ref{fact:reco_accross} states that charges on the two sides of a defect at the beginning of an iteration cannot recombine together during said iteration.

\begin{fact}[Charge between defects]\label{fact:between_defects}
    For all $t \geq 0$ and $\sigma \in \Sigma^{(t)}$ we have
    \begin{equation}
        Q_\sigma^{(t)} = 0. \label{eq:noflux}
    \end{equation}
\end{fact}

Finally, Fact~\ref{fact:between_defects} states that the total charge between two defects is always zero. Note that this also directly imply that the total charge in left open interval between two defects is always null. The proof of Lemma~\ref{lemma:ifp_3_s} follows directly from Fact~\ref{fact:conservation}: every positive charge can be paired up with a corresponding negative charge on its left. Since stack increments eventually transforms into anti-signals in the absence of defects, and since anti-signals propagates faster than forward-signals, all excitations will eventually recombine. Facts~\ref{fact:reco_accross} and \ref{fact:between_defects} will be used in the proof of Lemmas~\ref{lemma:ifp_2_s} and \ref{lemma:ifp_3_s}. We give the proof of all facts here.

\begin{proof}[Proof of Fact \ref{fact:conservation}]
\ref{eq:global} follows directly from the creation and recombination rules that always include both a $+1$ and a $-1$ charge. \ref{eq:deficit} follows from the fact that $+1$ and $-1$ charges are created on the same site, but a $-1$ charge cannot bypass a $+1$ charge to its right, because anti-signals recombine with any $+1$ charge they encounter, and a $+1$ charge cannot bypass a $-1$ charge to its left, because backward-signals recombine with any $-1$ charge they encounter.
\end{proof}

\begin{proof}[Proof of Facts~\ref{fact:reco_accross} and \ref{fact:between_defects}] Fact~\ref{fact:between_defects} is true at time $t = 0$. 
We prove that if Fact~\ref{fact:between_defects} holds at time $t$ then Fact~\ref{fact:reco_accross} is true for iteration $t$ (Step 1), and that if Fact~\ref{fact:between_defects} holds at time $t$ and Fact~\ref{fact:reco_accross} is true for iteration $t$ then Fact~\ref{fact:between_defects} holds at time $t+1$ (Step 2), finishing the proof by immediate recursion.

\textit{Step 1}. We start by proving that if Fact~\ref{fact:between_defects} holds at time $t$ then Fact~\ref{fact:reco_accross} holds for iteration $t$, i.e.\ here we assume that the total charge between two defects was null at time $t$, and we want to prove that charges do not recombine accross defects. Let $\sigma_a^{(t)} < \sigma_b^{(t)} \in \Sigma^{(t)}$ and $\sigma_a^{(t+1)} < \sigma_b^{(t+1)} \in \Sigma^{(t+1)}$ be a pair of successive defects respectively at time $t$ and $t+1$. Note that such identification is not ambiguous because the automaton either recombines defects by pair or displaces defects by one to the left. All types of charge recombinations across a defect during iteration $t$ in chronological order are

\begin{itemize}[leftmargin=1cm, align=right]
    \item[($i$)] A backward-signal from $[\sigma_a^{(t)}, \sigma_b^{(t)})$ recombining with an anti-signal from $\mathbb{Z}_{<  \sigma_a}^{(t)}$.
    \item[($ii$)] A backward-signal from $[\sigma_a^{(t)}, \sigma_b^{(t)})$ recombining with a stack increment from $\mathbb{Z}_{<  \sigma_a}^{(t)}$.
    \item[($iii$)] An anti-signal from $[\sigma_a^{(t)}, \sigma_b^{(t)})$ recombining with a forward-signal from $\mathbb{Z}_{\geq \sigma_b}^{(t)}$.
\end{itemize}

We omit the reciprocal event for each recombination type, i.e.\ the analogous process occurring in the adjacent defect interval, since such cases are symmetric and follow by identical arguments. It is clear that since Fact~\ref{fact:between_defects} holds at time $t$, Fact~\ref{fact:conservation} implies that a backward-signal propagating to the left will encounter a negative charge from $[\sigma_a^{(t)}, \sigma_b^{(t)})$ before it can exit. Since negative charges cannot move to the left and since the automaton checks after each backward-signals displacement whether recombination with negative charges are possible, the backward-signal will recombine before exiting $[\sigma_a^{(t)}, \sigma_b^{(t)})$ and no recombination with a negative charge from $\mathbb{Z}_{< \sigma_a}^{(t)}$ is possible. We have ruled out events ($i$) and ($ii$) during iteration $t$. Since a positive charge within $[\sigma_a^{(t)}, \sigma_b^{(t)})$ will always remain on the left of a forward-signal from $\mathbb{Z}_{\geq \sigma_b}^{(t)}$, a similar argument rules out event ($iii$) and finishes the proof of Fact~\ref{fact:reco_accross} for iteration $t$.

\medbreak

\textit{Step 2}. Now let us assume that Fact~\ref{fact:between_defects} holds at time $t$ and that Fact~\ref{fact:reco_accross} is true for iteration $t$, i.e.\ the total charge between two defects was null at time $t$ and no charge recombination have occurred across the defects during iteration $t$, and we want to prove that the total charge between two defects remains null at time $t+1$. In the absence of recombination across defects, the remaining possible violations of Fact~\ref{fact:between_defects} at time $t+1$ correspond to charges from $[\sigma_a^{(t)}, \sigma_b^{(t)})$ outside of $[\sigma_a^{(t+1)}, \sigma_b^{(t+1)})$ at time $t+1$. Omitting reciprocal events similarly as in Step 1, the four possible cases are

\begin{itemize}[leftmargin=1cm, align=right]
    \item[($i$)] A backward-signal from $[\sigma_a^{(t)}, \sigma_b^{(t)})$ in $\mathbb{Z}_{< \sigma_a}^{(t+1)}$ at time $t+1$.
    \item[($ii$)] A forward-signal from $[\sigma_a^{(t)}, \sigma_b^{(t)})$ in $\mathbb{Z}_{\geq \sigma_b}^{(t+1)}$ at time $t+1$.
    \item[($iii$)] An anti-signal from $[\sigma_a^{(t)}, \sigma_b^{(t)})$ in $\mathbb{Z}_{\geq \sigma_b}^{(t+1)}$ at time $t+1$.
    \item[($iv$)] A stack from $[\sigma_a^{(t)}, \sigma_b^{(t)})$ in $\mathbb{Z}_{\geq \sigma_b}^{(t+1)}$ at time $t+1$.
\end{itemize}

First, notice that no mechanism allows a negative charge (which can only move to the right) to bypass a defect (which can only move to the left) to its left within an iteration. Since in addition because of Fact~\ref{fact:reco_accross} a negative charge from $[\sigma_a^{(t)}, \sigma_b^{(t)})$ cannot have recombined with a positive charge from $\mathbb{Z}_{< \sigma_a}^{(t)}$, a backward-signal will always recombine before bypassing a defect to its left which forbids ($i$). This implies that the site of a defect and the site on its left are without backward-signals. Consequently, forward-signals encountering a defect always transform into backward-signals, which forbids ($ii$). A corollary of forbidding ($ii$) is that all positive charges between two successive defects remain on the left of the right defect. Therefore, the same argument as for ($i$) but in the other direction rules out cases ($iii$) and ($iv$). This finishes the proof.
\end{proof}

\subsection{Recombination of all excitations after correction}\label{appendix:recombination}

This section is devoted to the proof of Lemma \ref{lemma:ifp_3_s} that states that in a defect-free configuration, the successive application of the ASR ensures that all excitations eventually recombine. The key ingredient in the proof is  Fact~\ref{fact:conservation} that states that the global charge within the decoder is conserved and equal to zero, and that every positive charge can be paired with a negative charge on its left. In this situation, the higher speed of anti-signals ensures that all signals and stacks recombine in the end, which gives Lemma \ref{lemma:ifp_3_s}.

\begin{proof}[Proof of Lemma~\ref{lemma:ifp_3_s}]
Let $t_r \geq 0$, $z \in \mathbb{Z}$ and $\delta \geq 0$ be such that $\Sigma^{(t_r)}=\varnothing$ and $\mathrm{supp}(u^{(t_r)}) \subset [z,z+\delta]$: no new charge can be created at this point. Because the positive charge on a given site is bounded by $2$ (one forward-signal and one backward-signal) the total positive charge of the decoder at time $t_r$ is upper bounded by $2(\delta+1)$, and Fact~\ref{fact:conservation} directly implies that the total number of negative charges is upper bounded by the same quantity. This negative charge is distributed between stacks and anti-signals.

For $t \geq t_r$, we consider the rightmost site with a non-empty stack. At least one anti-signal leaves this site at each iteration, since either the site is occupied by an anti-signal and the stack cannot decrement, or the stack decrements by $1$ and creates an anti-signal propagating to the right. The cumulative sum of this quantity from iteration $t_r$ on is however ultimately bounded by the number of negative charges within the decoder at time $t_r$, previously upper bounded by $2(\delta+1)$. This directly implies that all stacks are necessarily empty at time $t_i=t_r+2(\delta+1)$. It is now left to upper bound the time of the last recombination between the last positive charge and the last anti-signal. In the worst case, this corresponds to an anti-signal located on site $z$ at time $t_i$, and the rightmost forward-signal on site $z+\delta$ at time $t_r$. Since anti-signal propagates by 3 at each iteration the recombination occurs at time $t_s$ such that

\begin{equation}
    3(t_s-t_i) = \delta + (t_s-t_r),
\end{equation}
which  gives $t_s - t_r \leq \lceil  (7\delta+6)/2 \rceil \leq 5\delta$ if $\delta \geq 3$, and it can be checked independently that the bound still works if $\delta \leq 2$. During this time the right most forward-signal has propagated by at most $5\delta$, this finishes the proof.
\end{proof}

\subsection{Defect recombination on the left of the interaction frontier}\label{appendix:independent_matching}

This section is devoted to the proof of Lemma~\ref{lemma:ifp_2_s} that states that on the left of the interaction frontier, an open interval between two successive odd and even defects in that order is filled with forward-signals but without negative charges (stack increments or anti-signals), while the open interval in the converse order is without forward-signals. Since an even defect is also separated by a distance of at least two with the next odd defect, and since forward-signals propagate by 1 at each iteration, the two defects do not interact. This means that even defects are displaced by 1 to the left at each iteration, until they recombine with the odd defect on its left that remains immobile. The operation takes a time equal to the distance between the two defects. The Lemma can be well understood graphically from Figure~\ref{figure:signal_rule} (a) but the proof is quite technical and may be skipped on a first reading. In the proof, we use two facts on the interaction frontier that are proven in Subsection~\ref{appendix:proof_of_facts}.

\medbreak

\begin{proof}[Proof of Lemma~\ref{lemma:ifp_2_s}]
The proposition is true for $t=0$ where we have $\varphi^{(0)}=\mathrm{min}(\Sigma)$. Let us assume that Lemma~\ref{lemma:ifp_2_s} is true up to some time $t \geq 0$, i.e.\ we have

\begin{align}
    \refstepcounter{equation}
    &\big(\Sigma^{(t)}_{12} \cap \mathbb{Z}_{\leq \varphi}^{(t)}\big) \subset \Phi_F^{(t)}, \tag*{(\theequation)($t$)}\label{eq:lemma2_1_t}\\
    \refstepcounter{equation}
    &\big(\Sigma^{(t)}_{12} \cap \Phi_N^{(t)} \cap \mathbb{Z}_{\leq \varphi}^{(t)}\big) = \varnothing, \tag*{(\theequation)($t$)}\label{eq:lemma2_2_t}\\
    \refstepcounter{equation}
    &\big(\Sigma^{(t)}_{21} \cap \Phi_F^{(t)} \cap \mathbb{Z}_{< \varphi}^{(t)}\big) = \varnothing. \tag*{(\theequation)($t$)}\label{eq:lemma2_3_t}
\end{align}

and 

\begin{align}
    \refstepcounter{equation}
    \forall \sigma_a \in \Sigma_2^{(t)}, \sigma_b &\in \Sigma_1^{(t)}, \; \text{if } \sigma_a < \sigma_b \leq \varphi^{(t)},\notag \\ & \text{ then } \sigma_b - \sigma_a \geq 2. \tag*{(\theequation)($t$)}\label{eq:lemma2_4_t}
\end{align}

We want to prove that Lemma~\ref{lemma:ifp_2_s} still holds at time $t+1$, i.e.\ (A19)($t+1$), (A20)($t+1$), (A21)($t+1$) and (A22)($t+1$). We decompose the proof by characterizing first the interval $\mathbb{Z}_{\leq \varphi}^{(t)}$ in Step 1, and the intervals $(\varphi^{(t)},\varphi^{(t+1)}]$ and $[\varphi^{(t)},\varphi^{(t+1)})$ in Step 2.

\medbreak

\textit{Step 1.} First, let us argue that we still have at time $t+1$

\setcounter{equation}{18}

\begin{align}
    \refstepcounter{equation}
    &\big(\Sigma^{(t+1)}_{12} \cap \mathbb{Z}_{\leq \varphi}^{(t)}\big) \subset \Phi_F^{(t+1)}, \tag*{(\theequation*)($t+1$)}\label{eq:lemma2_1*_t}\\
    \refstepcounter{equation}
    &\big(\Sigma^{(t+1)}_{12} \cap \Phi_N^{(t+1)} \cap \mathbb{Z}_{\leq \varphi}^{(t)}\big) = \varnothing, \tag*{(\theequation*)($t+1$)}\label{eq:lemma2_2*_t} \\
    \refstepcounter{equation}
    &\big(\Sigma^{(t+1)}_{21} \cap \Phi_F^{(t+1)} \cap \mathbb{Z}_{< \varphi}^{(t)}\big) = \varnothing, \tag*{(\theequation*)($t+1$)}\label{eq:lemma2_3*_t}
\end{align}

Notice that proving the Lemma would require $Z_{\leq \varphi}^{(t+1)}$ (resp. $Z_{< \varphi}^{(t+1)}$) instead of $Z_{\leq \varphi}^{(t)}$ (resp. $Z_{< \varphi}^{(t)}$). We also need

\begin{align}
    \refstepcounter{equation}
    \forall \sigma_a \in \Sigma_2^{(t+1)}, \sigma_b &\in \Sigma_1^{(t+1)}, \; \text{if } \sigma_a < \sigma_b \leq \varphi^{(t)},\notag \\ & \text{ then } \sigma_b - \sigma_a \geq 2. \tag*{(\theequation*)($t+1$)}\label{eq:lemma2_4*_t}
\end{align}

We start by proving \ref{eq:lemma2_2*_t}. Note first that the open interval was without negative charge at time $t$ and since charges do no propagate across defects because of Fact~\ref{fact:between_defects} it is only left to verify that no negative charge on the left defects site enters the open interval. Since the left neighbouring sites of odd defects on the left of the interaction frontier are without defect or forward-signal at time $t$, either odd defects remain immobile and the associated stack does not decrement, or they recombine with a defect on its right in which case the stack is necessarily empty. In the two cases we have \ref{eq:lemma2_2*_t}.

Now we turn to \ref{eq:lemma2_1*_t} and \ref{eq:lemma2_3*_t}. Since forward-signals propagate by $1$ at each iteration between two defects it is sufficient to consider what happens at the creation of the forward-signal on some defect $\Sigma^{(t)} \ni \sigma < \varphi^{(t)}$: the forward-signal should be erased if the defect is even, and it should keep propagating if the defect is odd. Intuitively, for isolated defects, even defects have a forward-signal on their left that induce the defect displacement, the associated stack then decrements into an anti-signal that erases the previously created forward-signal. Odd defects however remain immobile and forward-signals it created keep propagating. We formally discuss all cases in the following, including defects adjacent to each other. The two main cases are again:

\begin{equation*}
    D_1 : \sigma \in \Sigma_1^{(t)}, \qquad
    D_2 : \sigma \in \Sigma_2^{(t)}.
\end{equation*}

The subdivision of the case is as follows

\begin{itemize}[leftmargin=0.8cm, align=right]
    \item[$D_1$] Here $\Sigma_1^{(t)} \ni \sigma \leq \varphi^{(t)}$. \ref{eq:lemma2_4_t} implies that $\sigma - 1 \not \in \Sigma^{(t)}$: hence either $\sigma + 1 \not \in \Sigma^{(t)} \cup \Phi^{(t)}$, and the defect remains immobile and emits a forward-signal, or $\sigma + 1 \in \Sigma^{(t)}$ and the two defects recombine.
    \item[$D_2$] Here $\Sigma_2^{(t)} \ni \sigma \leq \varphi^{(t)}$. \ref{eq:lemma2_1_t} implies that $\sigma - 1 \in \Sigma^{(t)} \cup \Phi_F^{(t)}$. We distinguish between the cases ($i$) $\sigma + 1 \not \in \Sigma^{(t)}$ and ($ii$) $\sigma + 1 \in \Sigma^{(t)}$. If ($i$), either $\sigma-1 \in \Phi_F^{(t)} \setminus \Sigma^{(t)}$ and the defect $\sigma$ is displaced to the left, or $\sigma-1 \in \Sigma^{(t)}$ and \ref{eq:lemma2_4_t} implies that $\sigma - 2 \not \in \Sigma^{(t)}$, so that the two defects $\sigma$ and $\sigma-1$ recombine. If ($ii$), either $\sigma-1 \in \Phi_F^{(t)} \setminus \Sigma^{(t)}$ and the two defects $\sigma$ and $\sigma+1$ recombine, or $\sigma-1 \in \Sigma^{(t)}$ which reduces to the subcase of ($i$) discussed above. In all cases an emitted forward-signal is erased by the anti-signal created from the stack decrement.
\end{itemize}

Note that because of \ref{eq:lemma2_4_t} two adjacent defects recombining are necessarily an odd and an even defect in that order. In this case the left-open interval is the singleton $\{\sigma_a\}$ that is without forward-signal and backward-signal, which implies the absence of negative charge by Fact \ref{fact:between_defects}. The recombination of the two defect then leaves the corresponding sites empty which verifies \ref{eq:lemma2_3*_t}.

It is left to prove \ref{eq:lemma2_4*_t}. Since odd defects $\sigma_b \leq \varphi^{(t)}$ remain immobile unless they recombine, the only possible decrease of a distance between successive even and odd defects concerns an odd defect moving from $\varphi^{(t)}+1$ at time $t$ to $\varphi^{(t)}$ at time $t+1$. This violates \ref{eq:lemma2_4*_t} only if $\varphi^{(t)}-1$ is also occupied by a defect at that time. The latter defect was either ($i$) on site $\varphi^{(t)}-1$ at time $t$, or ($ii$) on site $\varphi^{(t)}$ at time $t$. ($i$) is not possible because the even defect would have been displaced to the left, and if ($ii$) the two defects would have recombined unless $\varphi^{(t)}-1$ was occupied by a defect at time $t$, which is forbidden by Fact~\ref{fact:nogo}. All cases lead to a contradiction, this finishes the proof of \ref{eq:lemma2_4*_t}.

\begin{fact}[Impossible case]\label{fact:nogo}
    Let $t \geq 0$ and $\varphi^{(t)} \in \mathbb{Z}$ be the interaction frontier at time $t$, then $\{\varphi^{(t)}-1,\varphi^{(t)},\varphi^{(t)}+1\} \not\subset \Sigma^{(t)}$.
\end{fact}

\medbreak
\textit{Step 2.} It is now left to extend the result to the incremented region on the left of the interaction frontier, i.e.\ the interval between $\varphi^{(t)}$ and $\varphi^{(t+1)}$, i.e.\

\setcounter{equation}{18}

\begin{align}
    \refstepcounter{equation}
    &\big(\Sigma^{(t+1)}_{12} \cap (\varphi^{(t)},\varphi^{(t+1)}]\big) \subset \Phi_F^{(t+1)}, \tag*{(\theequation$^\dagger$)($t+1$)}\label{eq:lemma2_1dag_t}\\
    \refstepcounter{equation}
    &\big(\Sigma^{(t+1)}_{12} \cap \Phi_N^{(t+1)} \cap (\varphi^{(t)},\varphi^{(t+1)}]\big) = \varnothing, \tag*{(\theequation$^\dagger$)($t+1$)}\label{eq:lemma2_2dag_t} \\
    \refstepcounter{equation}
    &\big(\Sigma^{(t+1)}_{21} \cap \Phi_F^{(t+1)} \cap [\varphi^{(t)},\varphi^{(t+1)})\big) = \varnothing. \tag*{(\theequation$^\dagger$)($t+1$)}\label{eq:lemma2_3dag_t}
\end{align}

With the convention that for $z \in \mathbb{Z}$, $[z,z) = (z,z] =\varnothing$. We also need

\begin{align}
    \refstepcounter{equation}
    \forall \sigma_a \in \Sigma_2^{(t+1)}, \sigma_b &\in \Sigma_1^{(t+1)}, \; \text{if } \sigma_a < \sigma_b \leq \varphi^{(t+1)},\notag \\ & \text{ then } \sigma_b - \sigma_a \geq 2. \tag*{(\theequation$^\dagger$)($t+1$)}\label{eq:lemma2_4dag_t}
\end{align}

We start by verifying that \ref{eq:lemma2_1dag_t}, \ref{eq:lemma2_2dag_t} and \ref{eq:lemma2_3dag_t} hold for the three cases corresponding to the three possible updates of $\varphi^{(t)}$ to $\varphi^{(t+1)}$. Intuitively, the proof can be understood as follows. If $A$, the interaction frontier either follows the free propagation of a forward-signal or the frontier reaches a defect. In both cases, the previous position of the frontier is now filled by a forward-signal. If $\neg A \land B$, the incremented region is empty and all desired properties are trivially true. Finally, $\neg A \land \neg B$ corresponds to the recombination of the forward-signal to which was attached the interaction frontier in the interval between an even and an odd defect. In that case the interaction frontier is updated to the next defect or forward-signal so that the right-open incremented region is without such excitations.

We formally treat all cases in the following. Each case is subdivided into at most four cases corresponding to the position of $\varphi^{(t)}$ with respect to odd and even defects, using following notations

\begin{align*}
    &F_1 : \varphi^{(t)} \in \Sigma_1^{(t)}, \qquad
    F_2 : \varphi^{(t)} \in \Sigma_2^{(t)}, \\
    &F_{12} : \varphi^{(t)} \in \Sigma_{12}^{(t)}, \; \; \; \; \; \;
    F_{21} : \varphi^{(t)} \in \Sigma_{21}^{(t)}.
\end{align*}

Since $B = (P_1 \text{ or } P_{12})$, we obtain the following subdivision of cases:

\begin{itemize}[leftmargin=1.2cm, align=right]
    \item[$A\qquad$] \begin{itemize}[leftmargin=0.7cm, align=right]
        \item[$F_1\qquad$] Here $(\varphi^{(t)},\varphi^{(t+1)}] = \{\varphi^{(t+1)}\}$, with $\varphi^{(t)} \in \Sigma_1^{(t)}$ and $\varphi^{(t+1)} = \varphi^{(t)} +1 \in \Sigma^{(t+1)} \cup \Phi_F^{(t+1)}$. We distinguish between ($i$) $\varphi^{(t+1)} \in \Sigma^{(t+1)}$ and ($ii$) $\varphi^{(t+1)} \in \Phi_F^{(t+1)} \setminus \Sigma^{(t+1)}$ with $\varphi^{(t+1)} \in \Sigma_{12}^{(t+1)}$. If ($i$), we have $(\varphi^{(t)},\varphi^{(t+1)}] \cap \Sigma_{12}^{(t+1)} = \varnothing$. If ($ii$), we have necessarily $\varphi^{(t+1)} \not\in \Phi_N^{(t+1)}$ otherwise the forward-signal would have recombined, hence $[\varphi^{(t)},\varphi^{(t+1)}) \cap \Phi_N^{(t+1)} = \varnothing$. In both cases, we have \ref{eq:lemma2_1dag_t} and \ref{eq:lemma2_2dag_t}. Finally, recall that $[\varphi^{(t)},\varphi^{(t+1)}) = \{\varphi^{(t)}\}$ with $\varphi^{(t)} \in \Sigma_1^{(t)}$. Because of \ref{eq:lemma2_3_t} and \ref{eq:lemma2_4_t}, we have $\varphi^{(t)}-1 \not \in \Sigma^{(t)} \cup \Phi_F^{(t)}$, hence $\varphi^{(t)} \not \in \Phi_F^{(t+1)}$ and \ref{eq:lemma2_3dag_t}.
        
        \item[$F_2\qquad$] Here $(\varphi^{(t)},\varphi^{(t+1)}] = \{\varphi^{(t+1)}\}$, with $\varphi^{(t)} \in \Sigma_2^{(t)}$ and $\varphi^{(t+1)} = \varphi^{(t)} +1 \in \Sigma^{(t+1)} \cup \Phi_F^{(t+1)}$. This means that $(\varphi^{(t)},\varphi^{(t+1)}] \cap \Sigma_{12}^{(t+1)} = \varnothing$ and we directly have \ref{eq:lemma2_1dag_t} and \ref{eq:lemma2_2dag_t}. Finally, recall that $[\varphi^{(t)},\varphi^{(t+1)}) = \{\varphi^{(t)}\}$ with $\varphi^{(t)} \in \Sigma_2^{(t)}$, because of \ref{eq:lemma2_1_t} the defect recombines or is displaced to the left at the next iteration, hence $\varphi^{(t)} \not \in \Phi_F^{(t+1)}$ and we have \ref{eq:lemma2_3dag_t}.
        
        \item[$F_{12}\qquad$] Here $(\varphi^{(t)},\varphi^{(t+1)}] = \{\varphi^{(t+1)}\}$ with $\varphi^{(t)} \in \Sigma_{12}^{(t)}$ and $\varphi^{(t+1)} = \varphi^{(t)} +1 \in \Sigma^{(t+1)} \cup \Phi_F^{(t+1)}$. The same reasoning used in case $F_1$ gives \ref{eq:lemma2_1dag_t} and \ref{eq:lemma2_2dag_t}. Finally, recall that $[\varphi^{(t)},\varphi^{(t+1)}) = \{\varphi^{(t)}\}$ with $\varphi^{(t)} \in \Sigma_{12}^{(t)}$, since defects are displaced by at most 1 per iteration, we have $\varphi^{(t)} \not \in \Sigma_{21}^{(t+1)}$ and \ref{eq:lemma2_3dag_t}.
        
        \item[$F_{21}\qquad$] Here $(\varphi^{(t)},\varphi^{(t+1)}] = \{\varphi^{(t+1)}\}$ with $\varphi^{(t)} \in \Sigma_{21}^{(t)}$ and $\varphi^{(t+1)} = \varphi^{(t)} +1 \in \Sigma^{(t+1)} \cup \Phi_F^{(t+1)}$. The same reasoning used in case $F_2$ gives \ref{eq:lemma2_1dag_t} and \ref{eq:lemma2_2dag_t}. Finally, recall that $[\varphi^{(t)},\varphi^{(t+1)}) = \{\varphi^{(t)}\}$ with $\varphi^{(t)} \in \Sigma_{21}^{(t)}$, because of \ref{eq:lemma2_3_t} we have either ($i$) $\varphi^{(t)}-1 \in \Sigma_2^{(t)}$ or ($ii$) $\varphi^{(t)}-1 \not \in \Sigma^{(t)} \cup \Phi_F^{(t)}$. If ($i$) the defect recombines, or is displaced to the left at the next iteration and the emitted forward-signal is erased during the iteration, which reduces to ($ii$) where we clearly have $\varphi^{(t)} \not \in \Phi_F^{(t+1)}$, hence \ref{eq:lemma2_3dag_t}.
    \end{itemize}
    \item[$\neg A \land B\qquad$] \begin{itemize}[leftmargin=0.7cm, align=right]
        \item[$F_1\qquad$] Here $[\varphi^{(t)},\varphi^{(t+1)}) = (\varphi^{(t)},\varphi^{(t+1)}] = \varnothing$ and we directly have \ref{eq:lemma2_1dag_t}, \ref{eq:lemma2_2dag_t} and \ref{eq:lemma2_3dag_t}.\\
        \item[$F_{12}\qquad$] Here $[\varphi^{(t)},\varphi^{(t+1)}) = (\varphi^{(t)},\varphi^{(t+1)}] = \varnothing$ and we directly have \ref{eq:lemma2_1dag_t}, \ref{eq:lemma2_2dag_t} and \ref{eq:lemma2_3dag_t}.
    \end{itemize}
    \item[$\neg A \land \neg B\qquad$] \begin{itemize}[leftmargin=0.7cm, align=right]
        \item[$F_2\qquad$] Here $(\varphi^{(t)},\varphi^{(t+1)}) \subset \Sigma_{21}^{(t+1)}$ and $\varphi^{(t+1)} \in \Sigma^{(t+1)} \cup \Phi_F^{(t+1)}$ so that $(\varphi^{(t)},\varphi^{(t+1)}] \subset \Sigma_{21}^{(t+1)} \cup \Sigma_{1}^{(t+1)}$, hence \ref{eq:lemma2_1dag_t} and \ref{eq:lemma2_2dag_t} . In addition because of \ref{eq:lemma2_1_t} and \ref{eq:lemma2_2_t} the left defect recombines, or is displaced to the left and the emitted forward-signal is erased so that $[\varphi^{(t)},\varphi^{(t+1)}) \cap \Phi_F^{(t+1)} = \varnothing$, hence \ref{eq:lemma2_3dag_t}.
        \item[$F_{21}\qquad$] The case reduces to $F_2$.
    \end{itemize}
\end{itemize}

It is now left to prove \ref{eq:lemma2_4dag_t}. Since we have already proven \ref{eq:lemma2_4*_t}, it is left to consider the case $\sigma_a \in \Sigma_2^{(t+1)}, \sigma_b \in \Sigma_1^{(t+1)}$, with $\sigma_a < \sigma_b \leq \varphi^{(t+1)}$ and $\varphi^{(t)} < \sigma_b$. Here because of Fact \ref{fact:onedefect} we have $\varphi^{(t+1)}=\sigma_b$. We also have $\sigma_a \leq \varphi^{(t)}$ since otherwise applying Fact \ref{fact:onedefect} with $\sigma_a$ and some other defect $\sigma < \sigma_a$ (always existing because $\sigma_a$ even) would imply $\sigma_a = \varphi^{(t+1)} = \sigma_b$ in contradiction with the initial assumption.

\begin{fact}[One defect at a time]\label{fact:onedefect}
    Let $\sigma^{(t)} \in \Sigma^{(t)}$ and $\sigma^{(t+1)} \in \Sigma^{(t+1)}$ be the position of a defect at times $t$ and $t+1$, such that $\varphi^{(t)} < \sigma^{(t)}$. If $\sigma^{(t+1)} \leq \varphi^{(t+1)}$, then $\varphi^{(t+1)} = \sigma^{(t+1)}$.
\end{fact}

Let us suppose, for the sake of contradiction, that $\sigma_b-\sigma_a=\varphi^{(t+1)}-\varphi^{(t)}=1$. In this case because defects can only move by at most 1 to the left per iteration, possible cases are restricted to

\begin{itemize}[leftmargin=2.2cm, align=right]
    \item[($i$)] $\Sigma^{(t)} \cap [\varphi^{(t)},\varphi^{(t)}+2] = \{\varphi^{(t)},\varphi^{(t)}+1\}$
    \item[($ii$)] $\Sigma^{(t)} \cap [\varphi^{(t)},\varphi^{(t)}+2] = \{\varphi^{(t)},\varphi^{(t)}+2\}$,
    \item[($iii$)] $\Sigma^{(t)} \cap [\varphi^{(t)},\varphi^{(t)}+2] = \{\varphi^{(t)}+1,\varphi^{(t)}+2\}$,
    \item[($iv$)] $\Sigma^{(t)} \cap [\varphi^{(t)},\varphi^{(t)}+2] = [\varphi^{(t)},\varphi^{(t)}+2]$.
\end{itemize}

If ($i$) or ($iv$) the two left defects would have recombined unless $\varphi^{(t)}-1 \in \Sigma^{(t)}$ which contradicts Fact \ref{fact:nogo}. If ($ii$) the left defect would have been displaced to the left or recombined with a defect on the left because of \ref{eq:lemma2_1_t}. If ($iii$) the two left defects would have recombined. All cases lead to a contradiction, hence $\sigma_b-\sigma_a \geq 2$. We have finished the proof of (A19)($t+1$), (A20)($t+1$), (A21)($t+1$) and (A22)($t+1$). Lemma~\ref{lemma:ifp_2_s} follows by recursion.
\end{proof}

\subsection{From one defect to the other}\label{appendix:from_one_defect}

This section is devoted to the proof of Lemma~\ref{lemma:ifp_1_s} that states that the interaction frontier goes from one defect to the next one in linear time in the distance traveled. Since defects can only move to the left this directly implies that the frontier eventually reaches the last defect in linear time with the initial error width.

\medbreak

\begin{proof}[Proof of Lemma~\ref{lemma:ifp_1_s}]
Let $t_1 \geq 0$ be such that $\varphi^{(t_1)} \in \Sigma^{(t_1)}$ and $\Sigma^{(t_1)} \cap \mathbb{Z}_{>\varphi}^{(t_1)} \neq \varnothing$. If $\varphi^{(t_1)} \in \Sigma_2^{(t_1)}$, $\varphi^{(t)}$ is trivially strictly increasing until $t_2 > t_1$ such that $\Sigma^{(t_2)} \cap \mathbb{Z}_{>\varphi}^{(t_2)} = \varnothing$ or $\varphi^{(t_2)} \in \Sigma_1^{(t_2)}$, with $t_2-t_1 \leq \sigma_{2m} - \varphi^{(t_1)}$ and $\varphi^{(t_2)}-\varphi^{(t_1)} \geq t_2-t_1$. In the following we consider the remaining case, $\varphi^{(t_1)} \in \Sigma_1^{(t_1)}$. Let us first prove that the following set is non-empty

\begin{align}
    \mathcal{T}_2 &= \{ t \geq t_1 \; | \; |\Sigma^{(t)} \cap \mathbb{Z}_{\leq \varphi}^{(t)}| \equiv 0 \pmod{2} \} \notag\\ &=\{ t\geq t_1 \; | \;  \neg B\}. \label{eq:bascule}
\end{align}

The interaction frontier increases by $1$ or remains constant at each iteration as long as we remain in the setting $|\Sigma^{(t)} \cap \mathbb{Z}_{\leq \varphi}^{(t)}| \equiv 1 \pmod{2}$. Since $\varphi^{(t)} \in \Sigma^{(t)} \cup \Phi_F^{(t)}$, either the forward-signal propagates or the defect emits a forward-signal and $\varphi^{(t+1)} = \varphi^{(t)}+1$, unless a recombination event occurred. However, the left defect is immobile because its left neighbouring site is without forward-signals or defects, see Lemma~\ref{lemma:ifp_2_s}, this implies that no new negative charge is created in the open interval $(\sigma_1,\varphi^{(t)}]$ which upper bounds the number of recombination events. The total number of defects being even, if $\Sigma^{(t)} \neq \varnothing$ there will always be at least one defect on the right of $\varphi^{(t)}$ and this defect cannot move to the right. Hence $\mathcal{T}_2$ is non-empty and we set $t_2 := \min(\mathcal{T}_2)$. It is left to upper-bound $t_2-t_1$. If $\Sigma^{(t_2)} = \varnothing$, because of defects parity necessarily the two defects $\varphi^{(t_2-1)}, \varphi^{(t_2-1)}+1 \in \Sigma^{(t_2-1)}$ recombined and in this case $t_2-t_1 = 1$ and $\varphi^{(t_2)}-\varphi^{(t_1)} \geq 1$. On the other hand, if $\Sigma^{(t_2)} \neq \varnothing$ and $|\Sigma^{(t_2)} \cap \mathbb{Z}_{\leq \varphi}^{(t_2)}| \equiv 0 \pmod{2}$, because of Fact~\ref{fact:onedefect} we have $\varphi^{(t_2)} \in \Sigma^{(t_2)}$: we consider this case in the following.

We will now study how this defect located at $\varphi^{(t_2)}$ at time $t_2$ evolves with time. We denote by $\sigma^{(t)}$ its location at time $t$, so that $\sigma^{t_2} = \varphi^{(t_2)}$. Let us assume first for simplicity that $(\varphi^{(t_1)},\sigma^{(t_1)}) \cap \Sigma^{(t_1)} = \varnothing$. In this situation no new negative charge is created within the open interval $(\varphi^{(t_1)},\sigma^{(t)})$ for $t_1 \leq t \leq t_2$, and the total negative charge within the interval is initially upper bounded by $2(\sigma^{(t_1)}-\varphi^{(t_1)})$ at time $t_1$. Since the quantity also bounds the number of non-increasing steps, this gives the following upper bound on $t_2-t_1$

\begin{align}
    t_2-t_1 <  (\sigma^{(t_1)}-\varphi^{(t_1)}) + &2(\sigma^{(t_1)}-\varphi^{(t_1)}) \notag \\ &= 3(\sigma^{(t_1)}-\varphi^{(t_1)}). \label{eq:if_bound_1}
\end{align}

Note that because $\sigma^{(t_1)} \leq \sigma_{2m}$ we also have $t_2-t_1 \leq 3(\sigma_{2m}-\varphi^{(t_1)})$. The last displacement of the right defect $\sigma$ before time $t_2$ is induced in the worst case by the last forward-signal emitted by the left defect at time $t_1-1$. The right defect displacement is upper bounded by $\lceil(\sigma^{(t_1)}-\varphi^{(t_1)})/2 \rceil$ between $t_1$ and $t_2$, which gives the following lower bound on $\varphi^{(t_2)}-\varphi^{(t_1)}$

\begin{equation}\label{eq:if_bound_2}
    \varphi^{(t_2)}-\varphi^{(t_1)} \geq \lfloor \frac{\sigma^{(t_1)}-\varphi^{(t_1)}}{2} \rfloor.
\end{equation}

Combining with with \eqref{eq:if_bound_1} and checking independently that $\varphi^{(t_2)}-\varphi^{(t_1)} \geq 1$, we obtain the desired inequality

\begin{align}
    \varphi^{(t_2)}-\varphi^{(t_1)} &\geq \mathrm{max}(\lfloor (t_2 - t_1)/6 \rfloor,1) \notag \\ &\geq (t_2-t_1)/11, \label{eq:if_bound_f}
\end{align}
where we used that $t_2-t_1$ is an integer.

Recall that we treated the simple case where the interval $(\varphi^{(t_1)},\sigma^{(t_1)})$ was without defect. In the general case however, if $(\varphi^{(t_1)},\sigma^{(t_1)}) \cap \Sigma^{(t_1)} \neq \varnothing$, necessarily those defects will recombine by pair before they would be reached by the interaction frontier. Let us consider the smallest $t_i > t_1$ such that $(\varphi^{(t_i)},\sigma^{(t_i)}) \cap \Sigma^{(t_i)} = \varnothing$ with $\sigma^{(t_i)} \in \Sigma^{(t_i)}$ the defect $\sigma$ at time $t_i$. Notice first that the previous reasoning can be applied similarly between $t_1$ and $t_i$. For iterations $t > t_i$, recombination events concern negative charges in the interval $[\varphi^{(t_i)},\sigma^{(t_i)})$ at time $t_i$. Since the total charge of the interval is positive at that time, by direct combination of Facts~\ref{fact:conservation} and \ref{fact:between_defects}, the previous reasoning is still valid and we have \eqref{eq:if_bound_f}.
\end{proof}

\subsection{Additional facts on the interaction frontier}\label{appendix:proof_of_facts}

We now detail the proof of some Facts we used in the proof of Lemma~\ref{lemma:ifp_2_s}.

\begin{proof}[Proof of Fact \ref{fact:nogo}]
Suppose, for the sake of contradiction, that using identical assumption we have

\begin{equation}
    \refstepcounter{equation}
    \{\varphi^{(t)}-1,\varphi^{(t)},\varphi^{(t)}+1\} \subset \Sigma^{(t)} \tag*{(\theequation)($t$)}\label{eq:fact4}
\end{equation}

Since in the code-capacity model no new defects are created, and since defects can only be displaced by at most $1$ to the left, the only possible configurations at time $t-1$ can be grouped as follows 

\begin{itemize}[leftmargin=0.7cm, align=right]
    \item[($i$)] $\Sigma^{(t-1)} \cap [\varphi^{(t)}-1,\varphi^{(t)}+2] = \{\varphi^{(t)},\varphi^{(t)}+1,\varphi^{(t)}+2\}$,
    \item[($ii$)] $\Sigma^{(t-1)} \cap [\varphi^{(t)}-1,\varphi^{(t)}+2] = \{\varphi^{(t)}-1,\varphi^{(t)}+1,\varphi^{(t)}+2\}$,
    \item[($iii$)] $\Sigma^{(t-1)} \cap [\varphi^{(t)}-1,\varphi^{(t)}+2] = \{\varphi^{(t)}-1,\varphi^{(t)},\varphi^{(t)}+1\}$,
    \item[($iv$)] $\Sigma^{(t-1)} \cap [\varphi^{(t)}-1,\varphi^{(t)}+2] = \{\varphi^{(t)}-1,\varphi^{(t)},\varphi^{(t)}+2\}$,
    \item[($v$)] $\Sigma^{(t-1)} \cap [\varphi^{(t)}-1,\varphi^{(t)}+2] = [\varphi^{(t)}-1,\varphi^{(t)}+2]$.
\end{itemize}

Because of the pre-processing correction step in the update rule, two defects recombine in cases ($i$) and ($ii$) which is not compatible with \ref{eq:fact4}. The two left defects also recombine in cases ($iii$), ($iv$) and ($v$) unless $\varphi^{(t)}-2 \in \Sigma^{(t-1)}$. In this case however, because of the update of the interaction frontier, we get $\{\varphi^{(t-1)}-1,\varphi^{(t-1)},\varphi^{(t-1)}+1\} \subset \Sigma^{(t-1)}$, i.e.\ (A27)($t-1)$. It is clear that repeating the argument until $t=0$ yields $\varphi^{(0)} \neq \min(\Sigma)$, in contradiction with the initialization of the interaction frontier.
\end{proof}

\begin{proof}[Proof of Fact \ref{fact:onedefect}]
We distinguish between the three cases of update of $\varphi^{(t)}$ to $\varphi^{(t+1)}$. If $A$, then necessarily $\varphi^{(t+1)} = \varphi^{(t)}+1 \in \Sigma^{(t+1)} \cup \Phi_F^{(t+1)}$ and because defects are displaced by at most 1 per iteration, we have either $\varphi^{(t+1)} = \sigma^{(t+1)}$ or $\varphi^{(t+1)} = \sigma^{(t+1)} +1$. In the latter case, the defect was displaced to the left and necessarily its previous site $\sigma^{(t+1)} +1$ is without forward-signal or defect, in contradiction $\sigma^{(t+1)}+1 \not \in \Sigma^{(t+1)} \cup \Phi_F^{(t+1)}$. Hence $\varphi^{(t+1)} = \sigma^{(t+1)}$ is the only possible case left. If $\neg A \land B$, then $\varphi^{(t+1)} = \varphi^{(t)}$ which is clearly incompatible with the assumptions. If $\neg A \land \neg B$ then because $\sigma^{(t+1)} \in \Sigma^{(t+1)} \cup \Phi_F^{(t+1)} \cap \mathbb{Z}_{> \varphi}^{(t)}$, necessarily $\varphi^{(t+1)} \leq \sigma^{(t+1)}$, hence $\varphi^{(t+1)} = \sigma^{(t+1)}$. This finishes the proof.
\end{proof}

\section{Proof of threshold}\label{appendix:proof_of_threshold}

The section is devoted to the proof of the code-capacity threshold theorem of the ASR decoder on a periodic lattice of size $n$. We consider the decoding to be successful if the initial error is corrected and if all variables of the decoder have returned to zero, conversely, a logical flip or a non zero decoder configuration is considered a logical error. We formally state a more precise version of Theorem~\ref{theorem:threshold_code_capacity} here.

\renewcommand{\thetheorem}{1$^*$}
\begin{theorem}[ASR code-capacity threshold]\label{theorem:threshold_code_capacity_s}
    Consider a family of 1D periodic lattices of size $n$. There exists $\varepsilon_{th}>0, \alpha_* >0$ and $\tau = \mathcal{O}(n)$ such that for $\varepsilon < \varepsilon_{\text{th}}$, the logical error rate $\varepsilon_L$ of the ASR applied for $\tau$ time steps to an initial error where each qubit is flipped independently and identically with probability $\varepsilon$ satisfies $\varepsilon_L \leq \mathcal{O}(\exp{[\frac{1}{2}n^{\alpha_*} \log{\varepsilon/\varepsilon_{th}}]})$.
\end{theorem}



The proof gives the following lower bounds $\varepsilon_{th} > 0.4\%$ and $\alpha_* > 0.12$, that are likely very conservative as indicated by numerical simulations in the stronger phenomenological model. The proof is inspired by previous work \cite{kubica2019cellular,bravyi2013quantum,harrington2004analysis,gacs2001reliable} but is simplified to take advantage of the single dimension of the system. An error configuration is decomposed into a hierarchy of connected components sufficiently far from each other such that a connected element of the lower level of the hierarchy is corrected independently from upper levels using Theorem \ref{theorem:erasure_s}. Logical errors then only arise in the presence of an element of the last level of the hierarchy. Counting the number of last-level representatives and bounding their weight gives Theorem~\ref{theorem:threshold_code_capacity_s}. Note that Theorem~\ref{theorem:threshold_code_capacity} follows directly from  Theorem~\ref{theorem:threshold_code_capacity_s} for any $\alpha < \alpha_*$.

\medbreak

\begin{proof}[Proof of Theorem~\ref{theorem:threshold_code_capacity_s}]
Let $E$ be the random variable over edges of $\mathbb{Z}_n$ such that $\mathbb{P}(E) = \varepsilon^{|E|}(1-\varepsilon)^{n-|E|}$. We consider $u^{(t)}$ for $t\geq 0$ the sequence of ASR configurations with initial error configuration $\Sigma^{(0)} = \partial E$.

\subsection{Hierarchical error decomposition}

Define $L > 0$ to be optimized later. We define a level-0 chunk $C_0$ to be an element of $E$, that it to say a single edge of $\mathbb{Z}_n$ corresponding to an error. A level-$k$ chunk $C_k=C_{k-1,1} \sqcup C_{k-1,2}$ is defined recursively to be the disjoint union of two level-($k-1$) chunks $C_{k-1,1}$ and $C_{k-1,2}$ such that $\mathrm{diam}(C_k) \leq L^k/2$. The level-$k$ error $E_k$ is defined to be the union of all level-$k$ chunks

\begin{equation}
    E_k = \bigcup_i C_{k,i}.
\end{equation}

By definition $E = E_0$, and we have the following sequence of inclusions for some $m >0$,

\begin{equation}
    E = E_0 \supseteq E_1 \supseteq ... \supseteq E_m \supsetneq E_{m+1} = \varnothing.
\end{equation}

We can define $F_k = E_{k-1} \setminus E_k$ up to the last level of the hierarchy so that we obtain a disjoint decomposition of $E$

\begin{equation}
    E = F_0 \sqcup ... \sqcup F_m.
\end{equation}

Each subset $F_k$ is furthermore decomposed into connected components that will be corrected independently. We say that a subset of errors $D_k \subset F_k$ is $\ell$-connected if it cannot be split into two disjoint non-empty sets $B_{k,1}$ and $B_{k,2}$ separated by more than $\ell$. That is to say, for any $B_{k,1}, B_{k,2} \neq \varnothing$, if $D_k = B_{k,1} \sqcup B_{k,2}$, then $d(B_{k,1},B_{k,2}) \leq \ell$. An $(\ell,k)$-connected component is a subset of $F_k$ that is $\ell$-connected and that is not strictly included in another $\ell$-connected subset of $F_k$. From the structure of the decomposition of $E$ into $F_k$ we can upper bound the size of connected components of $F_k$ as well as lower bound their distance from each other.

\begin{lemma}[Connected components \cite{kubica2019cellular,bravyi2013quantum}]\label{lemma:connected_components}
    Let $L \geq 6$ be some constant and a subset of errors $D_k \subseteq F_k$ be a $(L^k,k)$-connected component of $F_k$. Then, $\mathrm{diam}(D_k) \leq L^k$ and $d(D_k,E_k \setminus D_k) > L^{k+1}/3$.
\end{lemma}

We include the proof of Lemma \ref{lemma:connected_components} at the end of the section for completeness. Intuitively, having the diameter of connected components to be bounded and different connected components to be far from each other will enable a local decoder such as the ASR to erase them independently. Combining Theorem \ref{theorem:erasure_s} and Lemma \ref{lemma:connected_components} for the right choice of $\ell > 0$ then ensures that $(\ell,k)$-connected components are corrected independently.

\begin{lemma}[Hierarchical decoding]\label{lemma:hierarchical_decoding}
    Let $L \geq 232$, and $E$ be an error configuration that can be decomposed into $E = F_k \sqcup ... \sqcup F_{M-1}$ with $k \leq M-1$ for $M = \lfloor \log n / \log L \rfloor$. Consider $D_k$ a $(L^k,k)$-connected component of $F_k$, $D_k$ is corrected independently from $(F_k \setminus D_k) \sqcup F_{k+1} \sqcup ... \sqcup F_{M-1}$ in time $\tau_k \leq 77L^k$.
\end{lemma}

Here, independent correction means that no excitation or defect originating from this $(L^k,k)$-connected component meets any excitation or defect from another $(L^k,k)$-connected component. This implies that the presence of this connected component within the initial error configuration does not affect the decoder outcome. Successive applications of Lemma \ref{lemma:hierarchical_decoding} then directly show that a logical error is only possible in the presence of a level-$M$ chunk in the decomposition of $E$. It is left to upper bound this probability.

\begin{proof}[Proof of Lemma \ref{lemma:hierarchical_decoding}]
Consider an error $E$ with following hierarchical decomposition $E = F_k \sqcup ... \sqcup F_{M-1}$ with $k \leq M-1$ for $M = \lfloor \log n / \log L \rfloor$ and $D_k$ a $(L^k,k)$-connected component. Recall that we have $E = D_k \sqcup (E_k \setminus D_k)$.
 
We know that by Theorem~\ref{theorem:erasure_s} an isolated error $E' = D_k$ is corrected by $\tau_k < 77\textrm{diam}(D_k)$ successive application of the ASR with excitations propagating up to distance at most $77\textrm{diam}(D_k)$ to the right. Hence choosing $L$ such that $77\textrm{diam}(D_k) < d(D_k,E_k \setminus D_k)$ ensures that an excitation originating from another cluster on the left cannot reach $D_k$ before it is erased, and that excitations from $D_k$ cannot interfere with another cluster on the right. It suffices to replace the relevant diameters and distance by bounds from Lemma~\ref{lemma:connected_components} to obtain a sufficient condition on $L$

\begin{equation}
    77L^k < L^{k+1}/3,
\end{equation}

and that choosing $L = 232$ ensures $D_k$ is corrected independently from $E_k \setminus D_k$.
\end{proof}

\subsection{Probability of a level-M chunk}

Since for every $k \geq 1$ a level-$k$ chunk is composed of two disjoint level-$(k-1)$ chunks, level-$k$ chunks are of weight $2^k$ and their number of representatives up to translation, noted $N_k$, can be recursively upper bounded by

\begin{equation}
    N_{k+1} \leq N_k^2 \times L^k
\end{equation}

which initialized from $N_0 = 1$ gives for every $k \geq 1$

\begin{equation}
    N_{k} \leq L^{2^{k}-(k+1)}.
\end{equation}

Multiplying by $n$ to account for translated configurations, the logical error probability can be upper bounded by a quantity doubly exponential in $M$

\begin{equation}
    \varepsilon_L \leq nL^{-(M+1)}(L \varepsilon)^{2^{M}}.
\end{equation}

Using $M = \lfloor \log n / \log L \rfloor$ we retrieve Theorem~\ref{theorem:threshold_code_capacity_s} where $L=232$, $\varepsilon_{th} = 1/L > 0.4\%$, $\alpha_* = \log 2 / \log L > 0.12$ and $\tau = 77n/L$.
\end{proof}

\begin{proof}[Proof of Lemma \ref{lemma:connected_components}]
The proof by contradiction is taken from \cite{kubica2019cellular,bravyi2013quantum} and included for completeness. Let $D_k$ a be $(L^k,k)$-connected component of $F_k$. Suppose that either $\mathrm{diam}(D_k) > L^k$ or $d(D_k,E_k \setminus D_k) \leq L^{k+1}/3$. In the first case there exists $C_{0,1}$ and $C_{0,2} \in D_k$ two level-$0$ chunks such that $d(C_{0,1},C_{0,2}) > L^k$. Necessarily $C_{0,1}$ and $C_{0,2}$ belong to two disjoint $k$-level chunks $C_{k,1}$ and $C_{k,2}$. In addition, since $D_k$ is $L^k$-connected we can choose $C_{0,1}$ and $C_{0,2}$ such that $d(C_{0,1},C_{0,2}) \leq 2L^k$. By the triangle inequality we have for $L \geq 6$
    
\begin{align}
    \mathrm{diam}&(C_{k,1} \sqcup C_{k,2}) \notag \\ &\leq  \mathrm{diam}(C_{k,1}) + d(C_{k,1},C_{k,2}) + \mathrm{diam}(C_{k,2}) \notag \\ &\leq L^k/2 + 2L^k + L^k/2 \leq L^{k+1}/2.
\end{align}

This implies that $C_{k,1} \sqcup C_{k,2} \in E_{k+1}$ and subsequently $C_{0,1} \notin F_k$ which contradicts the initial assumption. In the latter case, i.e.\ if $d(D_k,E_k \setminus D_k) \leq L^{k+1}/3$, there exists $C_{0,1} \in D_k$ and $C_{0,2} \in E_k \setminus D_k$ such that $d(C_{0,1},C_{0,2}) \leq L^{k+1}/3$. Let the two $k$-level chunks $C_{k,1}$ and $C_{k,2}$ be such that $C_{0,1} \in C_{k,1}$ and $C_{0,2} \in C_{k,2}$. Note that necessarily $C_{k,1} \cap C_{k,2} = \varnothing$ otherwise $C_{k,1} \cup C_{k,2}$ is $L^k$-connected and $C_{0,2} \in D_k$. By the triangle inequality we have for $L \geq 6$
    
\begin{align}
    \mathrm{diam}&(C_{k,1} \sqcup C_{k,2}) \notag \\ &\leq  \mathrm{diam}(C_{k,1}) + d(C_{k,1},C_{k,2}) + \mathrm{diam}(C_{k,2}) \notag \\ &\leq L^k/2 + L^{k+1}/3 + L^k/2 \leq L^{k+1}/2,
\end{align}

which reduces to the contradiction of the former case.
\end{proof}

\section{Alternative proposals}\label{appendix:alternative_proposals}

\begin{figure*}
    \centering
    \includegraphics[width=\linewidth]{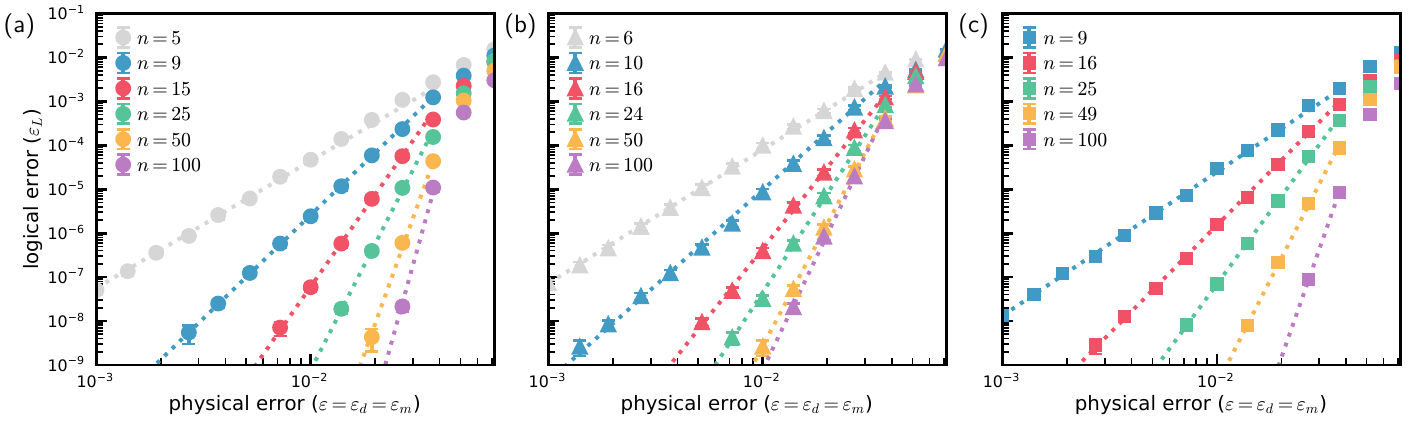}
    \caption{\label{figure:logical_alternative_proposals} Logical error rate as a function of the physical error rate for several system sizes in the phenomenological model for the (a) SSR decoder, (b) shearing-rule decoder and (c) Toom's rule decoder on a flat 2D surface. For each decoder, the data is fitted with the ansatz $An(B\varepsilon)^{\gamma_n}$ with a different parameter $\gamma_n$ for each $n$. $B^{-1}$ is estimated to $6.6\%$ ,$7.4\%$ and $7.7\%$ for the three decoders, respectively.}
\end{figure*}

The logical error rate of the shearing-rule and Toom's rule decoders is computed from Monte Carlo simulations, and plotted in Figure~\ref{figure:logical_alternative_proposals} along with a reproduction of Figure~\ref{figure:performances}~(a) for comparison with the SSR decoder. We assess the practical performance of the various decoders by fitting $\varepsilon_L$ with an ansatz of the form $An(B\varepsilon)^{\gamma_n}$, where the exponent $\gamma_n$ is allowed to depend on $n$ and is plotted in Figure~\ref{figure:performances} (c) of the main text.

\bibliography{bibliography}

\begin{thebibliography}{56}%
\makeatletter
\providecommand \@ifxundefined [1]{%
 \@ifx{#1\undefined}
}%
\providecommand \@ifnum [1]{%
 \ifnum #1\expandafter \@firstoftwo
 \else \expandafter \@secondoftwo
 \fi
}%
\providecommand \@ifx [1]{%
 \ifx #1\expandafter \@firstoftwo
 \else \expandafter \@secondoftwo
 \fi
}%
\providecommand \natexlab [1]{#1}%
\providecommand \enquote  [1]{``#1''}%
\providecommand \bibnamefont  [1]{#1}%
\providecommand \bibfnamefont [1]{#1}%
\providecommand \citenamefont [1]{#1}%
\providecommand \href@noop [0]{\@secondoftwo}%
\providecommand \href [0]{\begingroup \@sanitize@url \@href}%
\providecommand \@href[1]{\@@startlink{#1}\@@href}%
\providecommand \@@href[1]{\endgroup#1\@@endlink}%
\providecommand \@sanitize@url [0]{\catcode `\\12\catcode `\$12\catcode `\&12\catcode `\#12\catcode `\^12\catcode `\_12\catcode `\%12\relax}%
\providecommand \@@startlink[1]{}%
\providecommand \@@endlink[0]{}%
\providecommand \url  [0]{\begingroup\@sanitize@url \@url }%
\providecommand \@url [1]{\endgroup\@href {#1}{\urlprefix }}%
\providecommand \urlprefix  [0]{URL }%
\providecommand \Eprint [0]{\href }%
\providecommand \doibase [0]{https://doi.org/}%
\providecommand \selectlanguage [0]{\@gobble}%
\providecommand \bibinfo  [0]{\@secondoftwo}%
\providecommand \bibfield  [0]{\@secondoftwo}%
\providecommand \translation [1]{[#1]}%
\providecommand \BibitemOpen [0]{}%
\providecommand \bibitemStop [0]{}%
\providecommand \bibitemNoStop [0]{.\EOS\space}%
\providecommand \EOS [0]{\spacefactor3000\relax}%
\providecommand \BibitemShut  [1]{\csname bibitem#1\endcsname}%
\let\auto@bib@innerbib\@empty
\bibitem [{\citenamefont {Kitaev}(2003)}]{kitaev2003fault}%
  \BibitemOpen
  \bibfield  {author} {\bibinfo {author} {\bibfnamefont {A.~Y.}\ \bibnamefont {Kitaev}},\ }\bibfield  {title} {\bibinfo {title} {Fault-tolerant quantum computation by anyons},\ }\href {https://doi.org/10.1016/S0003-4916(02)00018-0} {\bibfield  {journal} {\bibinfo  {journal} {Annals of Physics}\ }\textbf {\bibinfo {volume} {303}},\ \bibinfo {pages} {2} (\bibinfo {year} {2003})}\BibitemShut {NoStop}%
\bibitem [{\citenamefont {Fowler}\ \emph {et~al.}(2012)\citenamefont {Fowler}, \citenamefont {Mariantoni}, \citenamefont {Martinis},\ and\ \citenamefont {Cleland}}]{fowler2012surface}%
  \BibitemOpen
  \bibfield  {author} {\bibinfo {author} {\bibfnamefont {A.~G.}\ \bibnamefont {Fowler}}, \bibinfo {author} {\bibfnamefont {M.}~\bibnamefont {Mariantoni}}, \bibinfo {author} {\bibfnamefont {J.~M.}\ \bibnamefont {Martinis}},\ and\ \bibinfo {author} {\bibfnamefont {A.~N.}\ \bibnamefont {Cleland}},\ }\bibfield  {title} {\bibinfo {title} {Surface codes: Towards practical large-scale quantum computation},\ }\href {https://doi.org/10.1103/PhysRevA.86.032324} {\bibfield  {journal} {\bibinfo  {journal} {Physical Review A}\ }\textbf {\bibinfo {volume} {86}},\ \bibinfo {pages} {032324} (\bibinfo {year} {2012})}\BibitemShut {NoStop}%
\bibitem [{goo(2023)}]{google2023suppressing}%
  \BibitemOpen
  \bibfield  {title} {\bibinfo {title} {Suppressing quantum errors by scaling a surface code logical qubit},\ }\href {https://doi.org/10.1038/s41586-022-05434-1} {\bibfield  {journal} {\bibinfo  {journal} {Nature}\ }\textbf {\bibinfo {volume} {614}},\ \bibinfo {pages} {676} (\bibinfo {year} {2023})}\BibitemShut {NoStop}%
\bibitem [{\citenamefont {Krinner}\ \emph {et~al.}(2022)\citenamefont {Krinner}, \citenamefont {Lacroix}, \citenamefont {Remm}, \citenamefont {Di~Paolo}, \citenamefont {Genois}, \citenamefont {Leroux}, \citenamefont {Hellings}, \citenamefont {Lazar}, \citenamefont {Swiadek}, \citenamefont {Herrmann} \emph {et~al.}}]{krinner2022realizing}%
  \BibitemOpen
  \bibfield  {author} {\bibinfo {author} {\bibfnamefont {S.}~\bibnamefont {Krinner}}, \bibinfo {author} {\bibfnamefont {N.}~\bibnamefont {Lacroix}}, \bibinfo {author} {\bibfnamefont {A.}~\bibnamefont {Remm}}, \bibinfo {author} {\bibfnamefont {A.}~\bibnamefont {Di~Paolo}}, \bibinfo {author} {\bibfnamefont {E.}~\bibnamefont {Genois}}, \bibinfo {author} {\bibfnamefont {C.}~\bibnamefont {Leroux}}, \bibinfo {author} {\bibfnamefont {C.}~\bibnamefont {Hellings}}, \bibinfo {author} {\bibfnamefont {S.}~\bibnamefont {Lazar}}, \bibinfo {author} {\bibfnamefont {F.}~\bibnamefont {Swiadek}}, \bibinfo {author} {\bibfnamefont {J.}~\bibnamefont {Herrmann}}, \emph {et~al.},\ }\bibfield  {title} {\bibinfo {title} {Realizing repeated quantum error correction in a distance-three surface code},\ }\href {https://doi.org/10.1038/s41586-022-04566-8} {\bibfield  {journal} {\bibinfo  {journal} {Nature}\ }\textbf {\bibinfo {volume} {605}},\ \bibinfo {pages} {669} (\bibinfo {year} {2022})}\BibitemShut {NoStop}%
\bibitem [{\citenamefont {Acharya}\ \emph {et~al.}(2024)\citenamefont {Acharya}, \citenamefont {Aghababaie-Beni}, \citenamefont {Aleiner}, \citenamefont {Andersen}, \citenamefont {Ansmann}, \citenamefont {Arute}, \citenamefont {Arya}, \citenamefont {Asfaw}, \citenamefont {Astrakhantsev}, \citenamefont {Atalaya} \emph {et~al.}}]{acharya2024quantum}%
  \BibitemOpen
  \bibfield  {author} {\bibinfo {author} {\bibfnamefont {R.}~\bibnamefont {Acharya}}, \bibinfo {author} {\bibfnamefont {L.}~\bibnamefont {Aghababaie-Beni}}, \bibinfo {author} {\bibfnamefont {I.}~\bibnamefont {Aleiner}}, \bibinfo {author} {\bibfnamefont {T.~I.}\ \bibnamefont {Andersen}}, \bibinfo {author} {\bibfnamefont {M.}~\bibnamefont {Ansmann}}, \bibinfo {author} {\bibfnamefont {F.}~\bibnamefont {Arute}}, \bibinfo {author} {\bibfnamefont {K.}~\bibnamefont {Arya}}, \bibinfo {author} {\bibfnamefont {A.}~\bibnamefont {Asfaw}}, \bibinfo {author} {\bibfnamefont {N.}~\bibnamefont {Astrakhantsev}}, \bibinfo {author} {\bibfnamefont {J.}~\bibnamefont {Atalaya}}, \emph {et~al.},\ }\bibfield  {title} {\bibinfo {title} {Quantum error correction below the surface code threshold},\ }\bibfield  {journal} {\bibinfo  {journal} {arXiv preprint arXiv:2408.13687}\ }\href {https://doi.org/10.48550/arXiv.2408.13687} {10.48550/arXiv.2408.13687} (\bibinfo {year} {2024})\BibitemShut {NoStop}%
\bibitem [{\citenamefont {Edmonds}(1965)}]{edmonds1965paths}%
  \BibitemOpen
  \bibfield  {author} {\bibinfo {author} {\bibfnamefont {J.}~\bibnamefont {Edmonds}},\ }\bibfield  {title} {\bibinfo {title} {Paths, trees, and flowers},\ }\href@noop {} {\bibfield  {journal} {\bibinfo  {journal} {Canadian Journal of mathematics}\ }\textbf {\bibinfo {volume} {17}},\ \bibinfo {pages} {449} (\bibinfo {year} {1965})}\BibitemShut {NoStop}%
\bibitem [{\citenamefont {Dennis}\ \emph {et~al.}(2002)\citenamefont {Dennis}, \citenamefont {Kitaev}, \citenamefont {Landahl},\ and\ \citenamefont {Preskill}}]{dennis2002topological}%
  \BibitemOpen
  \bibfield  {author} {\bibinfo {author} {\bibfnamefont {E.}~\bibnamefont {Dennis}}, \bibinfo {author} {\bibfnamefont {A.}~\bibnamefont {Kitaev}}, \bibinfo {author} {\bibfnamefont {A.}~\bibnamefont {Landahl}},\ and\ \bibinfo {author} {\bibfnamefont {J.}~\bibnamefont {Preskill}},\ }\bibfield  {title} {\bibinfo {title} {Topological quantum memory},\ }\href {https://doi.org/10.1063/1.1499754} {\bibfield  {journal} {\bibinfo  {journal} {Journal of Mathematical Physics}\ }\textbf {\bibinfo {volume} {43}},\ \bibinfo {pages} {4452} (\bibinfo {year} {2002})}\BibitemShut {NoStop}%
\bibitem [{\citenamefont {Delfosse}\ and\ \citenamefont {Nickerson}(2021)}]{delfosse2021almost}%
  \BibitemOpen
  \bibfield  {author} {\bibinfo {author} {\bibfnamefont {N.}~\bibnamefont {Delfosse}}\ and\ \bibinfo {author} {\bibfnamefont {N.~H.}\ \bibnamefont {Nickerson}},\ }\bibfield  {title} {\bibinfo {title} {Almost-linear time decoding algorithm for topological codes},\ }\href {https://doi.org/10.22331/q-2021-12-02-595} {\bibfield  {journal} {\bibinfo  {journal} {Quantum}\ }\textbf {\bibinfo {volume} {5}},\ \bibinfo {pages} {595} (\bibinfo {year} {2021})}\BibitemShut {NoStop}%
\bibitem [{\citenamefont {Chamberland}\ \emph {et~al.}(2020)\citenamefont {Chamberland}, \citenamefont {Kubica}, \citenamefont {Yoder},\ and\ \citenamefont {Zhu}}]{chamberland2020triangular}%
  \BibitemOpen
  \bibfield  {author} {\bibinfo {author} {\bibfnamefont {C.}~\bibnamefont {Chamberland}}, \bibinfo {author} {\bibfnamefont {A.}~\bibnamefont {Kubica}}, \bibinfo {author} {\bibfnamefont {T.~J.}\ \bibnamefont {Yoder}},\ and\ \bibinfo {author} {\bibfnamefont {G.}~\bibnamefont {Zhu}},\ }\bibfield  {title} {\bibinfo {title} {Triangular color codes on trivalent graphs with flag qubits},\ }\href {https://doi.org/10.1088/1367-2630/ab68fd} {\bibfield  {journal} {\bibinfo  {journal} {New Journal of Physics}\ }\textbf {\bibinfo {volume} {22}},\ \bibinfo {pages} {023019} (\bibinfo {year} {2020})}\BibitemShut {NoStop}%
\bibitem [{\citenamefont {Kubica}\ and\ \citenamefont {Delfosse}(2023)}]{kubica2023efficient}%
  \BibitemOpen
  \bibfield  {author} {\bibinfo {author} {\bibfnamefont {A.}~\bibnamefont {Kubica}}\ and\ \bibinfo {author} {\bibfnamefont {N.}~\bibnamefont {Delfosse}},\ }\bibfield  {title} {\bibinfo {title} {Efficient color code decoders in {$d \ge 2$} dimensions from toric code decoders},\ }\href {https://doi.org/10.22331/q-2023-02-21-929} {\bibfield  {journal} {\bibinfo  {journal} {Quantum}\ }\textbf {\bibinfo {volume} {7}},\ \bibinfo {pages} {929} (\bibinfo {year} {2023})}\BibitemShut {NoStop}%
\bibitem [{\citenamefont {Berent}\ \emph {et~al.}(2024)\citenamefont {Berent}, \citenamefont {Burgholzer}, \citenamefont {Derks}, \citenamefont {Eisert},\ and\ \citenamefont {Wille}}]{berent2024decoding}%
  \BibitemOpen
  \bibfield  {author} {\bibinfo {author} {\bibfnamefont {L.}~\bibnamefont {Berent}}, \bibinfo {author} {\bibfnamefont {L.}~\bibnamefont {Burgholzer}}, \bibinfo {author} {\bibfnamefont {P.-J.~H.}\ \bibnamefont {Derks}}, \bibinfo {author} {\bibfnamefont {J.}~\bibnamefont {Eisert}},\ and\ \bibinfo {author} {\bibfnamefont {R.}~\bibnamefont {Wille}},\ }\bibfield  {title} {\bibinfo {title} {Decoding quantum color codes with maxsat},\ }\href {https://doi.org/doi.org/10.22331/q-2024-10-23-1506} {\bibfield  {journal} {\bibinfo  {journal} {Quantum}\ }\textbf {\bibinfo {volume} {8}},\ \bibinfo {pages} {1506} (\bibinfo {year} {2024})}\BibitemShut {NoStop}%
\bibitem [{\citenamefont {Bomb{\'\i}n}(2015)}]{bombin2015single}%
  \BibitemOpen
  \bibfield  {author} {\bibinfo {author} {\bibfnamefont {H.}~\bibnamefont {Bomb{\'\i}n}},\ }\bibfield  {title} {\bibinfo {title} {Single-shot fault-tolerant quantum error correction},\ }\href {https://doi.org/10.1103/PhysRevX.5.031043} {\bibfield  {journal} {\bibinfo  {journal} {Physical Review X}\ }\textbf {\bibinfo {volume} {5}},\ \bibinfo {pages} {031043} (\bibinfo {year} {2015})}\BibitemShut {NoStop}%
\bibitem [{\citenamefont {Chesi}\ \emph {et~al.}(2010)\citenamefont {Chesi}, \citenamefont {R{\"o}thlisberger},\ and\ \citenamefont {Loss}}]{chesi2010self}%
  \BibitemOpen
  \bibfield  {author} {\bibinfo {author} {\bibfnamefont {S.}~\bibnamefont {Chesi}}, \bibinfo {author} {\bibfnamefont {B.}~\bibnamefont {R{\"o}thlisberger}},\ and\ \bibinfo {author} {\bibfnamefont {D.}~\bibnamefont {Loss}},\ }\bibfield  {title} {\bibinfo {title} {Self-correcting quantum memory in a thermal environment},\ }\href {https://doi.org/10.1103/PhysRevA.82.022305} {\bibfield  {journal} {\bibinfo  {journal} {Physical Review A}\ }\textbf {\bibinfo {volume} {82}},\ \bibinfo {pages} {022305} (\bibinfo {year} {2010})}\BibitemShut {NoStop}%
\bibitem [{\citenamefont {Yoshida}(2011)}]{yoshida2011feasibility}%
  \BibitemOpen
  \bibfield  {author} {\bibinfo {author} {\bibfnamefont {B.}~\bibnamefont {Yoshida}},\ }\bibfield  {title} {\bibinfo {title} {Feasibility of self-correcting quantum memory and thermal stability of topological order},\ }\href {https://doi.org/10.1016/j.aop.2011.06.001} {\bibfield  {journal} {\bibinfo  {journal} {Annals of Physics}\ }\textbf {\bibinfo {volume} {326}},\ \bibinfo {pages} {2566} (\bibinfo {year} {2011})}\BibitemShut {NoStop}%
\bibitem [{\citenamefont {Landon-Cardinal}\ \emph {et~al.}(2015)\citenamefont {Landon-Cardinal}, \citenamefont {Yoshida}, \citenamefont {Poulin},\ and\ \citenamefont {Preskill}}]{landon2015perturbative}%
  \BibitemOpen
  \bibfield  {author} {\bibinfo {author} {\bibfnamefont {O.}~\bibnamefont {Landon-Cardinal}}, \bibinfo {author} {\bibfnamefont {B.}~\bibnamefont {Yoshida}}, \bibinfo {author} {\bibfnamefont {D.}~\bibnamefont {Poulin}},\ and\ \bibinfo {author} {\bibfnamefont {J.}~\bibnamefont {Preskill}},\ }\bibfield  {title} {\bibinfo {title} {Perturbative instability of quantum memory based on effective long-range interactions},\ }\href {https://doi.org/10.1103/PhysRevA.91.032303} {\bibfield  {journal} {\bibinfo  {journal} {Physical Review A}\ }\textbf {\bibinfo {volume} {91}},\ \bibinfo {pages} {032303} (\bibinfo {year} {2015})}\BibitemShut {NoStop}%
\bibitem [{\citenamefont {Brown}\ \emph {et~al.}(2016)\citenamefont {Brown}, \citenamefont {Loss}, \citenamefont {Pachos}, \citenamefont {Self},\ and\ \citenamefont {Wootton}}]{brown2016quantum}%
  \BibitemOpen
  \bibfield  {author} {\bibinfo {author} {\bibfnamefont {B.~J.}\ \bibnamefont {Brown}}, \bibinfo {author} {\bibfnamefont {D.}~\bibnamefont {Loss}}, \bibinfo {author} {\bibfnamefont {J.~K.}\ \bibnamefont {Pachos}}, \bibinfo {author} {\bibfnamefont {C.~N.}\ \bibnamefont {Self}},\ and\ \bibinfo {author} {\bibfnamefont {J.~R.}\ \bibnamefont {Wootton}},\ }\bibfield  {title} {\bibinfo {title} {Quantum memories at finite temperature},\ }\href {https://doi.org/10.1103/RevModPhys.88.045005} {\bibfield  {journal} {\bibinfo  {journal} {Reviews of Modern Physics}\ }\textbf {\bibinfo {volume} {88}},\ \bibinfo {pages} {045005} (\bibinfo {year} {2016})}\BibitemShut {NoStop}%
\bibitem [{\citenamefont {Alicki}\ \emph {et~al.}(2010)\citenamefont {Alicki}, \citenamefont {Horodecki}, \citenamefont {Horodecki},\ and\ \citenamefont {Horodecki}}]{alicki2010thermal}%
  \BibitemOpen
  \bibfield  {author} {\bibinfo {author} {\bibfnamefont {R.}~\bibnamefont {Alicki}}, \bibinfo {author} {\bibfnamefont {M.}~\bibnamefont {Horodecki}}, \bibinfo {author} {\bibfnamefont {P.}~\bibnamefont {Horodecki}},\ and\ \bibinfo {author} {\bibfnamefont {R.}~\bibnamefont {Horodecki}},\ }\bibfield  {title} {\bibinfo {title} {On thermal stability of topological qubit in kitaev's 4d model},\ }\href {https://doi.org/10.48550/arXiv.0811.0033} {\bibfield  {journal} {\bibinfo  {journal} {Open Systems \& Information Dynamics}\ }\textbf {\bibinfo {volume} {17}},\ \bibinfo {pages} {1} (\bibinfo {year} {2010})}\BibitemShut {NoStop}%
\bibitem [{\citenamefont {Bravyi}\ and\ \citenamefont {Terhal}(2009)}]{bravyi2009no}%
  \BibitemOpen
  \bibfield  {author} {\bibinfo {author} {\bibfnamefont {S.}~\bibnamefont {Bravyi}}\ and\ \bibinfo {author} {\bibfnamefont {B.}~\bibnamefont {Terhal}},\ }\bibfield  {title} {\bibinfo {title} {A no-go theorem for a two-dimensional self-correcting quantum memory based on stabilizer codes},\ }\href {https://doi.org/10.1088/1367-2630/11/4/043029} {\bibfield  {journal} {\bibinfo  {journal} {New Journal of Physics}\ }\textbf {\bibinfo {volume} {11}},\ \bibinfo {pages} {043029} (\bibinfo {year} {2009})}\BibitemShut {NoStop}%
\bibitem [{\citenamefont {Bravyi}\ and\ \citenamefont {Haah}(2013)}]{bravyi2013quantum}%
  \BibitemOpen
  \bibfield  {author} {\bibinfo {author} {\bibfnamefont {S.}~\bibnamefont {Bravyi}}\ and\ \bibinfo {author} {\bibfnamefont {J.}~\bibnamefont {Haah}},\ }\bibfield  {title} {\bibinfo {title} {Quantum self-correction in the 3d cubic code model},\ }\href {https://doi.org/10.1103/PhysRevLett.111.200501} {\bibfield  {journal} {\bibinfo  {journal} {Physical Review Letters}\ }\textbf {\bibinfo {volume} {111}},\ \bibinfo {pages} {200501} (\bibinfo {year} {2013})}\BibitemShut {NoStop}%
\bibitem [{\citenamefont {Lin}\ \emph {et~al.}(2024)\citenamefont {Lin}, \citenamefont {Wang},\ and\ \citenamefont {Hsieh}}]{lin2024proposals}%
  \BibitemOpen
  \bibfield  {author} {\bibinfo {author} {\bibfnamefont {T.-C.}\ \bibnamefont {Lin}}, \bibinfo {author} {\bibfnamefont {H.-P.}\ \bibnamefont {Wang}},\ and\ \bibinfo {author} {\bibfnamefont {M.-H.}\ \bibnamefont {Hsieh}},\ }\bibfield  {title} {\bibinfo {title} {Proposals for 3d self-correcting quantum memory},\ }\href@noop {} {\bibfield  {journal} {\bibinfo  {journal} {arXiv preprint arXiv:2411.03115}\ } (\bibinfo {year} {2024})}\BibitemShut {NoStop}%
\bibitem [{\citenamefont {G{\'a}cs}(2001)}]{gacs2001reliable}%
  \BibitemOpen
  \bibfield  {author} {\bibinfo {author} {\bibfnamefont {P.}~\bibnamefont {G{\'a}cs}},\ }\bibfield  {title} {\bibinfo {title} {Reliable cellular automata with self-organization},\ }\href@noop {} {\bibfield  {journal} {\bibinfo  {journal} {Journal of Statistical Physics}\ }\textbf {\bibinfo {volume} {103}},\ \bibinfo {pages} {45} (\bibinfo {year} {2001})}\BibitemShut {NoStop}%
\bibitem [{\citenamefont {Cirel’son}(2006)}]{cirel2006reliable}%
  \BibitemOpen
  \bibfield  {author} {\bibinfo {author} {\bibfnamefont {B.}~\bibnamefont {Cirel’son}},\ }\bibfield  {title} {\bibinfo {title} {Reliable storage of information in a system of unreliable components with local interactions}\ }(\bibinfo {organization} {Springer},\ \bibinfo {year} {2006})\BibitemShut {NoStop}%
\bibitem [{\citenamefont {Balasubramanian}\ \emph {et~al.}(2024)\citenamefont {Balasubramanian}, \citenamefont {Davydova},\ and\ \citenamefont {Lake}}]{balasubramanian2024local}%
  \BibitemOpen
  \bibfield  {author} {\bibinfo {author} {\bibfnamefont {S.}~\bibnamefont {Balasubramanian}}, \bibinfo {author} {\bibfnamefont {M.}~\bibnamefont {Davydova}},\ and\ \bibinfo {author} {\bibfnamefont {E.}~\bibnamefont {Lake}},\ }\bibfield  {title} {\bibinfo {title} {A local automaton for the 2d toric code},\ }\bibfield  {journal} {\bibinfo  {journal} {arXiv preprint arXiv:2412.19803}\ }\href {https://doi.org/10.48550/arXiv.2412.19803} {10.48550/arXiv.2412.19803} (\bibinfo {year} {2024})\BibitemShut {NoStop}%
\bibitem [{\citenamefont {Harrington}(2004)}]{harrington2004analysis}%
  \BibitemOpen
  \bibfield  {author} {\bibinfo {author} {\bibfnamefont {J.~W.}\ \bibnamefont {Harrington}},\ }\emph {\bibinfo {title} {Analysis of quantum error-correcting codes: symplectic lattice codes and toric codes}},\ \href@noop {} {Ph.D. thesis},\ \bibinfo  {school} {California Institute of Technology} (\bibinfo {year} {2004})\BibitemShut {NoStop}%
\bibitem [{\citenamefont {Herold}\ \emph {et~al.}(2015)\citenamefont {Herold}, \citenamefont {Campbell}, \citenamefont {Eisert},\ and\ \citenamefont {Kastoryano}}]{herold2015cellular}%
  \BibitemOpen
  \bibfield  {author} {\bibinfo {author} {\bibfnamefont {M.}~\bibnamefont {Herold}}, \bibinfo {author} {\bibfnamefont {E.~T.}\ \bibnamefont {Campbell}}, \bibinfo {author} {\bibfnamefont {J.}~\bibnamefont {Eisert}},\ and\ \bibinfo {author} {\bibfnamefont {M.~J.}\ \bibnamefont {Kastoryano}},\ }\bibfield  {title} {\bibinfo {title} {Cellular-automaton decoders for topological quantum memories},\ }\href {https://doi.org/10.1038/npjqi.2015.10} {\bibfield  {journal} {\bibinfo  {journal} {npj Quantum information}\ }\textbf {\bibinfo {volume} {1}},\ \bibinfo {pages} {1} (\bibinfo {year} {2015})}\BibitemShut {NoStop}%
\bibitem [{\citenamefont {Herold}\ \emph {et~al.}(2017)\citenamefont {Herold}, \citenamefont {Kastoryano}, \citenamefont {Campbell},\ and\ \citenamefont {Eisert}}]{herold2017cellular}%
  \BibitemOpen
  \bibfield  {author} {\bibinfo {author} {\bibfnamefont {M.}~\bibnamefont {Herold}}, \bibinfo {author} {\bibfnamefont {M.~J.}\ \bibnamefont {Kastoryano}}, \bibinfo {author} {\bibfnamefont {E.~T.}\ \bibnamefont {Campbell}},\ and\ \bibinfo {author} {\bibfnamefont {J.}~\bibnamefont {Eisert}},\ }\bibfield  {title} {\bibinfo {title} {Cellular automaton decoders of topological quantum memories in the fault tolerant setting},\ }\href {https://doi.org/10.1088/1367-2630/aa7099} {\bibfield  {journal} {\bibinfo  {journal} {New Journal of Physics}\ }\textbf {\bibinfo {volume} {19}},\ \bibinfo {pages} {063012} (\bibinfo {year} {2017})}\BibitemShut {NoStop}%
\bibitem [{\citenamefont {Toom}(1995)}]{toom1995cellular}%
  \BibitemOpen
  \bibfield  {author} {\bibinfo {author} {\bibfnamefont {A.}~\bibnamefont {Toom}},\ }\bibfield  {title} {\bibinfo {title} {Cellular automata with errors: problems for students of probability},\ }\href@noop {} {\bibfield  {journal} {\bibinfo  {journal} {Topics in contemporary probability and its applications}\ ,\ \bibinfo {pages} {117}} (\bibinfo {year} {1995})}\BibitemShut {NoStop}%
\bibitem [{\citenamefont {Toom}(1980)}]{toom1980stable}%
  \BibitemOpen
  \bibfield  {author} {\bibinfo {author} {\bibfnamefont {A.~L.}\ \bibnamefont {Toom}},\ }\bibfield  {title} {\bibinfo {title} {Stable and attractive trajectories in multicomponent systems},\ }\href@noop {} {\bibfield  {journal} {\bibinfo  {journal} {Multicomponent random systems}\ }\textbf {\bibinfo {volume} {6}},\ \bibinfo {pages} {549} (\bibinfo {year} {1980})}\BibitemShut {NoStop}%
\bibitem [{\citenamefont {Kubica}\ and\ \citenamefont {Preskill}(2019)}]{kubica2019cellular}%
  \BibitemOpen
  \bibfield  {author} {\bibinfo {author} {\bibfnamefont {A.}~\bibnamefont {Kubica}}\ and\ \bibinfo {author} {\bibfnamefont {J.}~\bibnamefont {Preskill}},\ }\bibfield  {title} {\bibinfo {title} {Cellular-automaton decoders with provable thresholds for topological codes},\ }\href {https://doi.org/10.1103/PhysRevLett.123.020501} {\bibfield  {journal} {\bibinfo  {journal} {Physical Review Letters}\ }\textbf {\bibinfo {volume} {123}},\ \bibinfo {pages} {020501} (\bibinfo {year} {2019})}\BibitemShut {NoStop}%
\bibitem [{\citenamefont {Breuckmann}\ \emph {et~al.}(2016)\citenamefont {Breuckmann}, \citenamefont {Duivenvoorden}, \citenamefont {Michels},\ and\ \citenamefont {Terhal}}]{breuckmann2016local}%
  \BibitemOpen
  \bibfield  {author} {\bibinfo {author} {\bibfnamefont {N.~P.}\ \bibnamefont {Breuckmann}}, \bibinfo {author} {\bibfnamefont {K.}~\bibnamefont {Duivenvoorden}}, \bibinfo {author} {\bibfnamefont {D.}~\bibnamefont {Michels}},\ and\ \bibinfo {author} {\bibfnamefont {B.~M.}\ \bibnamefont {Terhal}},\ }\bibfield  {title} {\bibinfo {title} {Local decoders for the 2d and 4d toric code},\ }\bibfield  {journal} {\bibinfo  {journal} {arXiv preprint arXiv:1609.00510}\ }\href {https://doi.org/10.48550/arXiv.1609.00510} {10.48550/arXiv.1609.00510} (\bibinfo {year} {2016})\BibitemShut {NoStop}%
\bibitem [{\citenamefont {Guedes}\ \emph {et~al.}(2024)\citenamefont {Guedes}, \citenamefont {Winter},\ and\ \citenamefont {M{\"u}ller}}]{guedes2024quantum}%
  \BibitemOpen
  \bibfield  {author} {\bibinfo {author} {\bibfnamefont {T.~L.}\ \bibnamefont {Guedes}}, \bibinfo {author} {\bibfnamefont {D.}~\bibnamefont {Winter}},\ and\ \bibinfo {author} {\bibfnamefont {M.}~\bibnamefont {M{\"u}ller}},\ }\bibfield  {title} {\bibinfo {title} {Quantum cellular automata for quantum error correction and density classification},\ }\href {https://doi.org/10.1103/PhysRevLett.133.150601} {\bibfield  {journal} {\bibinfo  {journal} {Physical Review Letters}\ }\textbf {\bibinfo {volume} {133}},\ \bibinfo {pages} {150601} (\bibinfo {year} {2024})}\BibitemShut {NoStop}%
\bibitem [{\citenamefont {Vasmer}\ \emph {et~al.}(2021)\citenamefont {Vasmer}, \citenamefont {Browne},\ and\ \citenamefont {Kubica}}]{vasmer2021cellular}%
  \BibitemOpen
  \bibfield  {author} {\bibinfo {author} {\bibfnamefont {M.}~\bibnamefont {Vasmer}}, \bibinfo {author} {\bibfnamefont {D.~E.}\ \bibnamefont {Browne}},\ and\ \bibinfo {author} {\bibfnamefont {A.}~\bibnamefont {Kubica}},\ }\bibfield  {title} {\bibinfo {title} {Cellular automaton decoders for topological quantum codes with noisy measurements and beyond},\ }\href {https://doi.org/doi.org/10.1038/s41598-021-81138-2} {\bibfield  {journal} {\bibinfo  {journal} {Scientific reports}\ }\textbf {\bibinfo {volume} {11}},\ \bibinfo {pages} {2027} (\bibinfo {year} {2021})}\BibitemShut {NoStop}%
\bibitem [{\citenamefont {Ueno}\ \emph {et~al.}(2021)\citenamefont {Ueno}, \citenamefont {Kondo}, \citenamefont {Tanaka}, \citenamefont {Suzuki},\ and\ \citenamefont {Tabuchi}}]{ueno2021qecool}%
  \BibitemOpen
  \bibfield  {author} {\bibinfo {author} {\bibfnamefont {Y.}~\bibnamefont {Ueno}}, \bibinfo {author} {\bibfnamefont {M.}~\bibnamefont {Kondo}}, \bibinfo {author} {\bibfnamefont {M.}~\bibnamefont {Tanaka}}, \bibinfo {author} {\bibfnamefont {Y.}~\bibnamefont {Suzuki}},\ and\ \bibinfo {author} {\bibfnamefont {Y.}~\bibnamefont {Tabuchi}},\ }\bibfield  {title} {\bibinfo {title} {Qecool: On-line quantum error correction with a superconducting decoder for surface code},\ }in\ \href@noop {} {\emph {\bibinfo {booktitle} {2021 58th ACM/IEEE Design Automation Conference (DAC)}}}\ (\bibinfo {organization} {IEEE},\ \bibinfo {year} {2021})\ pp.\ \bibinfo {pages} {451--456}\BibitemShut {NoStop}%
\bibitem [{\citenamefont {Leghtas}\ \emph {et~al.}(2015)\citenamefont {Leghtas}, \citenamefont {Touzard}, \citenamefont {Pop}, \citenamefont {Kou}, \citenamefont {Vlastakis}, \citenamefont {Petrenko}, \citenamefont {Sliwa}, \citenamefont {Narla}, \citenamefont {Shankar}, \citenamefont {Hatridge} \emph {et~al.}}]{leghtas2015confining}%
  \BibitemOpen
  \bibfield  {author} {\bibinfo {author} {\bibfnamefont {Z.}~\bibnamefont {Leghtas}}, \bibinfo {author} {\bibfnamefont {S.}~\bibnamefont {Touzard}}, \bibinfo {author} {\bibfnamefont {I.~M.}\ \bibnamefont {Pop}}, \bibinfo {author} {\bibfnamefont {A.}~\bibnamefont {Kou}}, \bibinfo {author} {\bibfnamefont {B.}~\bibnamefont {Vlastakis}}, \bibinfo {author} {\bibfnamefont {A.}~\bibnamefont {Petrenko}}, \bibinfo {author} {\bibfnamefont {K.~M.}\ \bibnamefont {Sliwa}}, \bibinfo {author} {\bibfnamefont {A.}~\bibnamefont {Narla}}, \bibinfo {author} {\bibfnamefont {S.}~\bibnamefont {Shankar}}, \bibinfo {author} {\bibfnamefont {M.~J.}\ \bibnamefont {Hatridge}}, \emph {et~al.},\ }\bibfield  {title} {\bibinfo {title} {Confining the state of light to a quantum manifold by engineered two-photon loss},\ }\href {https://doi.org/10.1126/science.aaa20} {\bibfield  {journal} {\bibinfo  {journal} {Science}\ }\textbf {\bibinfo {volume} {347}},\ \bibinfo {pages} {853} (\bibinfo {year} {2015})}\BibitemShut {NoStop}%
\bibitem [{\citenamefont {Mirrahimi}\ \emph {et~al.}(2014)\citenamefont {Mirrahimi}, \citenamefont {Leghtas}, \citenamefont {Albert}, \citenamefont {Touzard}, \citenamefont {Schoelkopf}, \citenamefont {Jiang},\ and\ \citenamefont {Devoret}}]{mirrahimi2014dynamically}%
  \BibitemOpen
  \bibfield  {author} {\bibinfo {author} {\bibfnamefont {M.}~\bibnamefont {Mirrahimi}}, \bibinfo {author} {\bibfnamefont {Z.}~\bibnamefont {Leghtas}}, \bibinfo {author} {\bibfnamefont {V.~V.}\ \bibnamefont {Albert}}, \bibinfo {author} {\bibfnamefont {S.}~\bibnamefont {Touzard}}, \bibinfo {author} {\bibfnamefont {R.~J.}\ \bibnamefont {Schoelkopf}}, \bibinfo {author} {\bibfnamefont {L.}~\bibnamefont {Jiang}},\ and\ \bibinfo {author} {\bibfnamefont {M.~H.}\ \bibnamefont {Devoret}},\ }\bibfield  {title} {\bibinfo {title} {Dynamically protected cat-qubits: a new paradigm for universal quantum computation},\ }\href {https://doi.org/10.1088/1367-2630/16/4/045014} {\bibfield  {journal} {\bibinfo  {journal} {New Journal of Physics}\ }\textbf {\bibinfo {volume} {16}},\ \bibinfo {pages} {045014} (\bibinfo {year} {2014})}\BibitemShut {NoStop}%
\bibitem [{\citenamefont {Puri}\ \emph {et~al.}(2017)\citenamefont {Puri}, \citenamefont {Boutin},\ and\ \citenamefont {Blais}}]{puri2017engineering}%
  \BibitemOpen
  \bibfield  {author} {\bibinfo {author} {\bibfnamefont {S.}~\bibnamefont {Puri}}, \bibinfo {author} {\bibfnamefont {S.}~\bibnamefont {Boutin}},\ and\ \bibinfo {author} {\bibfnamefont {A.}~\bibnamefont {Blais}},\ }\bibfield  {title} {\bibinfo {title} {Engineering the quantum states of light in a kerr-nonlinear resonator by two-photon driving},\ }\href {https://doi.org/10.1038/s41534-017-0019-1} {\bibfield  {journal} {\bibinfo  {journal} {npj Quantum Information}\ }\textbf {\bibinfo {volume} {3}},\ \bibinfo {pages} {18} (\bibinfo {year} {2017})}\BibitemShut {NoStop}%
\bibitem [{\citenamefont {Guillaud}\ and\ \citenamefont {Mirrahimi}(2019)}]{guillaud2019repetition}%
  \BibitemOpen
  \bibfield  {author} {\bibinfo {author} {\bibfnamefont {J.}~\bibnamefont {Guillaud}}\ and\ \bibinfo {author} {\bibfnamefont {M.}~\bibnamefont {Mirrahimi}},\ }\bibfield  {title} {\bibinfo {title} {Repetition cat qubits for fault-tolerant quantum computation},\ }\href {https://doi.org/10.1103/PhysRevX.9.041053} {\bibfield  {journal} {\bibinfo  {journal} {Physical Review X}\ }\textbf {\bibinfo {volume} {9}},\ \bibinfo {pages} {041053} (\bibinfo {year} {2019})}\BibitemShut {NoStop}%
\bibitem [{\citenamefont {Ruiz}\ \emph {et~al.}(2024)\citenamefont {Ruiz}, \citenamefont {Guillaud}, \citenamefont {Leverrier}, \citenamefont {Mirrahimi},\ and\ \citenamefont {Vuillot}}]{ruiz2024ldpc}%
  \BibitemOpen
  \bibfield  {author} {\bibinfo {author} {\bibfnamefont {D.}~\bibnamefont {Ruiz}}, \bibinfo {author} {\bibfnamefont {J.}~\bibnamefont {Guillaud}}, \bibinfo {author} {\bibfnamefont {A.}~\bibnamefont {Leverrier}}, \bibinfo {author} {\bibfnamefont {M.}~\bibnamefont {Mirrahimi}},\ and\ \bibinfo {author} {\bibfnamefont {C.}~\bibnamefont {Vuillot}},\ }\bibfield  {title} {\bibinfo {title} {Ldpc-cat codes for low-overhead quantum computing in 2d},\ }\bibfield  {journal} {\bibinfo  {journal} {arXiv preprint arXiv:2401.09541}\ }\href {https://doi.org/10.48550/arXiv.2401.09541} {10.48550/arXiv.2401.09541} (\bibinfo {year} {2024})\BibitemShut {NoStop}%
\bibitem [{\citenamefont {Chamberland}\ \emph {et~al.}(2022)\citenamefont {Chamberland}, \citenamefont {Noh}, \citenamefont {Arrangoiz-Arriola}, \citenamefont {Campbell}, \citenamefont {Hann}, \citenamefont {Iverson}, \citenamefont {Putterman}, \citenamefont {Bohdanowicz}, \citenamefont {Flammia}, \citenamefont {Keller} \emph {et~al.}}]{chamberland2022building}%
  \BibitemOpen
  \bibfield  {author} {\bibinfo {author} {\bibfnamefont {C.}~\bibnamefont {Chamberland}}, \bibinfo {author} {\bibfnamefont {K.}~\bibnamefont {Noh}}, \bibinfo {author} {\bibfnamefont {P.}~\bibnamefont {Arrangoiz-Arriola}}, \bibinfo {author} {\bibfnamefont {E.~T.}\ \bibnamefont {Campbell}}, \bibinfo {author} {\bibfnamefont {C.~T.}\ \bibnamefont {Hann}}, \bibinfo {author} {\bibfnamefont {J.}~\bibnamefont {Iverson}}, \bibinfo {author} {\bibfnamefont {H.}~\bibnamefont {Putterman}}, \bibinfo {author} {\bibfnamefont {T.~C.}\ \bibnamefont {Bohdanowicz}}, \bibinfo {author} {\bibfnamefont {S.~T.}\ \bibnamefont {Flammia}}, \bibinfo {author} {\bibfnamefont {A.}~\bibnamefont {Keller}}, \emph {et~al.},\ }\bibfield  {title} {\bibinfo {title} {Building a fault-tolerant quantum computer using concatenated cat codes},\ }\href {https://doi.org/10.1103/PRXQuantum.3.010329} {\bibfield  {journal} {\bibinfo  {journal} {PRX Quantum}\ }\textbf {\bibinfo {volume} {3}},\ \bibinfo {pages} {010329} (\bibinfo {year} {2022})}\BibitemShut {NoStop}%
\bibitem [{\citenamefont {Putterman}\ \emph {et~al.}(2024)\citenamefont {Putterman}, \citenamefont {Noh}, \citenamefont {Hann}, \citenamefont {MacCabe}, \citenamefont {Aghaeimeibodi}, \citenamefont {Patel}, \citenamefont {Lee}, \citenamefont {Jones}, \citenamefont {Moradinejad}, \citenamefont {Rodriguez} \emph {et~al.}}]{putterman2024hardware}%
  \BibitemOpen
  \bibfield  {author} {\bibinfo {author} {\bibfnamefont {H.}~\bibnamefont {Putterman}}, \bibinfo {author} {\bibfnamefont {K.}~\bibnamefont {Noh}}, \bibinfo {author} {\bibfnamefont {C.~T.}\ \bibnamefont {Hann}}, \bibinfo {author} {\bibfnamefont {G.~S.}\ \bibnamefont {MacCabe}}, \bibinfo {author} {\bibfnamefont {S.}~\bibnamefont {Aghaeimeibodi}}, \bibinfo {author} {\bibfnamefont {R.~N.}\ \bibnamefont {Patel}}, \bibinfo {author} {\bibfnamefont {M.}~\bibnamefont {Lee}}, \bibinfo {author} {\bibfnamefont {W.~M.}\ \bibnamefont {Jones}}, \bibinfo {author} {\bibfnamefont {H.}~\bibnamefont {Moradinejad}}, \bibinfo {author} {\bibfnamefont {R.}~\bibnamefont {Rodriguez}}, \emph {et~al.},\ }\bibfield  {title} {\bibinfo {title} {Hardware-efficient quantum error correction using concatenated bosonic qubits},\ }\bibfield  {journal} {\bibinfo  {journal} {arXiv preprint arXiv:2409.13025}\ }\href {https://doi.org/10.48550/arXiv.2409.13025} {10.48550/arXiv.2409.13025} (\bibinfo {year} {2024})\BibitemShut {NoStop}%
\bibitem [{\citenamefont {Lieu}\ \emph {et~al.}(2024)\citenamefont {Lieu}, \citenamefont {Liu},\ and\ \citenamefont {Gorshkov}}]{lieu2024candidate}%
  \BibitemOpen
  \bibfield  {author} {\bibinfo {author} {\bibfnamefont {S.}~\bibnamefont {Lieu}}, \bibinfo {author} {\bibfnamefont {Y.-J.}\ \bibnamefont {Liu}},\ and\ \bibinfo {author} {\bibfnamefont {A.~V.}\ \bibnamefont {Gorshkov}},\ }\bibfield  {title} {\bibinfo {title} {Candidate for a passively protected quantum memory in two dimensions},\ }\href {https://doi.org/10.1103/PhysRevLett.133.030601} {\bibfield  {journal} {\bibinfo  {journal} {Physical Review Letters}\ }\textbf {\bibinfo {volume} {133}},\ \bibinfo {pages} {030601} (\bibinfo {year} {2024})}\BibitemShut {NoStop}%
\bibitem [{\citenamefont {Gottesman}(2013)}]{gottesman2013fault}%
  \BibitemOpen
  \bibfield  {author} {\bibinfo {author} {\bibfnamefont {D.}~\bibnamefont {Gottesman}},\ }\bibfield  {title} {\bibinfo {title} {Fault-tolerant quantum computation with constant overhead},\ }\bibfield  {journal} {\bibinfo  {journal} {arXiv preprint arXiv:1310.2984}\ }\href {https://doi.org/10.48550/arXiv.1310.2984} {10.48550/arXiv.1310.2984} (\bibinfo {year} {2013})\BibitemShut {NoStop}%
\bibitem [{\citenamefont {Michnicki}(2015)}]{michnicki2015towards}%
  \BibitemOpen
  \bibfield  {author} {\bibinfo {author} {\bibfnamefont {K.}~\bibnamefont {Michnicki}},\ }\href@noop {} {\emph {\bibinfo {title} {Towards self-correcting quantum memories}}}\ (\bibinfo  {publisher} {University of Washington},\ \bibinfo {year} {2015})\BibitemShut {NoStop}%
\bibitem [{\citenamefont {Lake}(2025)}]{lake2025fast}%
  \BibitemOpen
  \bibfield  {author} {\bibinfo {author} {\bibfnamefont {E.}~\bibnamefont {Lake}},\ }\bibfield  {title} {\bibinfo {title} {Fast offline decoding with local message-passing automata},\ }\bibfield  {journal} {\bibinfo  {journal} {arXiv preprint arXiv:2506.03266}\ }\href {https://doi.org/10.48550/arXiv.2506.03266} {10.48550/arXiv.2506.03266} (\bibinfo {year} {2025})\BibitemShut {NoStop}%
\bibitem [{\citenamefont {Lang}\ and\ \citenamefont {B{\"u}chler}(2018)}]{lang2018strictly}%
  \BibitemOpen
  \bibfield  {author} {\bibinfo {author} {\bibfnamefont {N.}~\bibnamefont {Lang}}\ and\ \bibinfo {author} {\bibfnamefont {H.~P.}\ \bibnamefont {B{\"u}chler}},\ }\bibfield  {title} {\bibinfo {title} {Strictly local one-dimensional topological quantum error correction with symmetry-constrained cellular automata},\ }\href {https://doi.org/10.21468/SciPostPhys.4.1.007} {\bibfield  {journal} {\bibinfo  {journal} {SciPost Physics}\ }\textbf {\bibinfo {volume} {4}},\ \bibinfo {pages} {007} (\bibinfo {year} {2018})}\BibitemShut {NoStop}%
\bibitem [{\citenamefont {G{\'a}cs}\ \emph {et~al.}(1978)\citenamefont {G{\'a}cs}, \citenamefont {Kurdyumov},\ and\ \citenamefont {Levin}}]{gacs1978one}%
  \BibitemOpen
  \bibfield  {author} {\bibinfo {author} {\bibfnamefont {P.}~\bibnamefont {G{\'a}cs}}, \bibinfo {author} {\bibfnamefont {G.~L.}\ \bibnamefont {Kurdyumov}},\ and\ \bibinfo {author} {\bibfnamefont {L.~A.}\ \bibnamefont {Levin}},\ }\bibfield  {title} {\bibinfo {title} {One-dimensional uniform arrays that wash out finite islands},\ }\href@noop {} {\bibfield  {journal} {\bibinfo  {journal} {Problemy Peredachi Informatsii}\ }\textbf {\bibinfo {volume} {14}},\ \bibinfo {pages} {92} (\bibinfo {year} {1978})}\BibitemShut {NoStop}%
\bibitem [{\citenamefont {Rozendaal}\ and\ \citenamefont {Z{\'e}mor}(2023)}]{rozendaal2023worst}%
  \BibitemOpen
  \bibfield  {author} {\bibinfo {author} {\bibfnamefont {W.}~\bibnamefont {Rozendaal}}\ and\ \bibinfo {author} {\bibfnamefont {G.}~\bibnamefont {Z{\'e}mor}},\ }\bibfield  {title} {\bibinfo {title} {A worst-case analysis of a renormalisation decoder for kitaev’s toric code}\ }(\bibinfo {organization} {IEEE},\ \bibinfo {year} {2023})\ pp.\ \bibinfo {pages} {625--629}\BibitemShut {NoStop}%
\bibitem [{\citenamefont {Wolf-Gladrow}(2004)}]{wolf2004lattice}%
  \BibitemOpen
  \bibfield  {author} {\bibinfo {author} {\bibfnamefont {D.~A.}\ \bibnamefont {Wolf-Gladrow}},\ }\href@noop {} {\emph {\bibinfo {title} {Lattice-gas cellular automata and lattice Boltzmann models: an introduction}}}\ (\bibinfo  {publisher} {Springer},\ \bibinfo {year} {2004})\BibitemShut {NoStop}%
\bibitem [{\citenamefont {Nagel}\ and\ \citenamefont {Schreckenberg}(1992)}]{nagel1992cellular}%
  \BibitemOpen
  \bibfield  {author} {\bibinfo {author} {\bibfnamefont {K.}~\bibnamefont {Nagel}}\ and\ \bibinfo {author} {\bibfnamefont {M.}~\bibnamefont {Schreckenberg}},\ }\bibfield  {title} {\bibinfo {title} {A cellular automaton model for freeway traffic},\ }\href@noop {} {\bibfield  {journal} {\bibinfo  {journal} {Journal de physique I}\ }\textbf {\bibinfo {volume} {2}},\ \bibinfo {pages} {2221} (\bibinfo {year} {1992})}\BibitemShut {NoStop}%
\bibitem [{\citenamefont {Hattori}\ and\ \citenamefont {Takesue}(1991)}]{hattori1991additive}%
  \BibitemOpen
  \bibfield  {author} {\bibinfo {author} {\bibfnamefont {T.}~\bibnamefont {Hattori}}\ and\ \bibinfo {author} {\bibfnamefont {S.}~\bibnamefont {Takesue}},\ }\bibfield  {title} {\bibinfo {title} {Additive conserved quantities in discrete-time lattice dynamical systems},\ }\href@noop {} {\bibfield  {journal} {\bibinfo  {journal} {Physica D: Nonlinear Phenomena}\ }\textbf {\bibinfo {volume} {49}},\ \bibinfo {pages} {295} (\bibinfo {year} {1991})}\BibitemShut {NoStop}%
\bibitem [{\citenamefont {Redeker}(2022)}]{redeker2022number}%
  \BibitemOpen
  \bibfield  {author} {\bibinfo {author} {\bibfnamefont {M.}~\bibnamefont {Redeker}},\ }\bibfield  {title} {\bibinfo {title} {Number conservation via particle flow in one-dimensional cellular automata},\ }\href@noop {} {\bibfield  {journal} {\bibinfo  {journal} {Fundamenta Informaticae}\ }\textbf {\bibinfo {volume} {187}} (\bibinfo {year} {2022})}\BibitemShut {NoStop}%
\bibitem [{\citenamefont {Wolnik}\ \emph {et~al.}(2020)\citenamefont {Wolnik}, \citenamefont {Nenca}, \citenamefont {Baetens},\ and\ \citenamefont {De~Baets}}]{wolnik2020split}%
  \BibitemOpen
  \bibfield  {author} {\bibinfo {author} {\bibfnamefont {B.}~\bibnamefont {Wolnik}}, \bibinfo {author} {\bibfnamefont {A.}~\bibnamefont {Nenca}}, \bibinfo {author} {\bibfnamefont {J.~M.}\ \bibnamefont {Baetens}},\ and\ \bibinfo {author} {\bibfnamefont {B.}~\bibnamefont {De~Baets}},\ }\bibfield  {title} {\bibinfo {title} {A split-and-perturb decomposition of number-conserving cellular automata},\ }\href@noop {} {\bibfield  {journal} {\bibinfo  {journal} {Physica D: Nonlinear Phenomena}\ }\textbf {\bibinfo {volume} {413}},\ \bibinfo {pages} {132645} (\bibinfo {year} {2020})}\BibitemShut {NoStop}%
\bibitem [{\citenamefont {Boccara}\ and\ \citenamefont {Fuks}(1998)}]{boccara1998cellular}%
  \BibitemOpen
  \bibfield  {author} {\bibinfo {author} {\bibfnamefont {N.}~\bibnamefont {Boccara}}\ and\ \bibinfo {author} {\bibfnamefont {H.}~\bibnamefont {Fuks}},\ }\bibfield  {title} {\bibinfo {title} {Cellular automaton rules conserving the number of active sites},\ }\href@noop {} {\bibfield  {journal} {\bibinfo  {journal} {Journal of Physics A: Mathematical and General}\ }\textbf {\bibinfo {volume} {31}},\ \bibinfo {pages} {6007} (\bibinfo {year} {1998})}\BibitemShut {NoStop}%
\bibitem [{\citenamefont {Boccara}\ and\ \citenamefont {Fuk{\'s}}(2002)}]{boccara2002number}%
  \BibitemOpen
  \bibfield  {author} {\bibinfo {author} {\bibfnamefont {N.}~\bibnamefont {Boccara}}\ and\ \bibinfo {author} {\bibfnamefont {H.}~\bibnamefont {Fuk{\'s}}},\ }\bibfield  {title} {\bibinfo {title} {Number-conserving cellular automaton rules},\ }\href@noop {} {\bibfield  {journal} {\bibinfo  {journal} {Fundamenta Informaticae}\ }\textbf {\bibinfo {volume} {52}},\ \bibinfo {pages} {1} (\bibinfo {year} {2002})}\BibitemShut {NoStop}%
\bibitem [{\citenamefont {Park}\ \emph {et~al.}(2024)\citenamefont {Park}, \citenamefont {Maskara}, \citenamefont {Kalinowski},\ and\ \citenamefont {Lukin}}]{park2024enhancing}%
  \BibitemOpen
  \bibfield  {author} {\bibinfo {author} {\bibfnamefont {M.}~\bibnamefont {Park}}, \bibinfo {author} {\bibfnamefont {N.}~\bibnamefont {Maskara}}, \bibinfo {author} {\bibfnamefont {M.}~\bibnamefont {Kalinowski}},\ and\ \bibinfo {author} {\bibfnamefont {M.~D.}\ \bibnamefont {Lukin}},\ }\bibfield  {title} {\bibinfo {title} {Enhancing quantum memory lifetime with measurement-free local error correction and reinforcement learning},\ }\bibfield  {journal} {\bibinfo  {journal} {arXiv preprint arXiv:2408.09524}\ }\href {https://doi.org/10.48550/arXiv.2408.09524} {10.48550/arXiv.2408.09524} (\bibinfo {year} {2024})\BibitemShut {NoStop}%
\bibitem [{git(2025)}]{git}%
  \BibitemOpen
  \href {https://doi.org/doi.org/10.5281/zenodo.17100661} {\bibinfo {title} {Code is available at \url{https://github.com/lpaletta/local-decoder}}} (\bibinfo {year} {2025})\BibitemShut {NoStop}%
\end{thebibliography}%

\end{document}